\documentclass[%
 aip,
 8pt,
 amsmath,amssymb,
 reprint,%
]{revtex4-2}

\usepackage{amsmath}
\usepackage{lineno}
\usepackage[colorlinks,linkcolor=red,citecolor=red,urlcolor=blue]{hyperref}

\usepackage{amssymb}
\usepackage{graphicx}
\usepackage{dcolumn}
\usepackage{bm}
\usepackage{color}
\usepackage{multirow}
\usepackage{subfigure}
\usepackage{verbatim}
\usepackage{tikz}
\usepackage{braket}
\usepackage[english]{babel}
\allowdisplaybreaks
\usepackage{cleveref}
\usepackage{nomencl}
\makenomenclature
\usepackage{array}
\newcolumntype{L}[1]{>{\raggedright\let\newline\\\arraybackslash\hspace{0pt}}m{#1}}
\newcolumntype{C}[1]{>{\centering\let\newline\\\arraybackslash\hspace{0pt}}m{#1}}
\newcolumntype{R}[1]{>{\raggedleft\let\newline\\\arraybackslash\hspace{0pt}}m{#1}}

\usepackage{etoolbox}
\renewcommand\nomgroup[1]{%
  \item[\bfseries
  \ifstrequal{#1}{A}{Abbreviations}{%
  \ifstrequal{#1}{B}{Dimensionless numbers}{
  \ifstrequal{#1}{C}{Superscripts}{%
  \ifstrequal{#1}{D}{Subscripts}{%
  \ifstrequal{#1}{E}{Symbols}{%
  }}}}}%
]}

\begin{document}

\title{A fluctuating lattice Boltzmann formulation based on orthogonal central moments}

\author{Alessandro De Rosis}
\affiliation{Department of Mechanical and Aerospace  Engineering, School of Engineering, The University of Manchester, Oxford Road, M13 9PL, Manchester, United Kingdom}
\author{Yang Zhou}
\affiliation{Department of Mechanical and Aerospace  Engineering, School of Engineering, The University of Manchester, Oxford Road, M13 9PL, Manchester, United Kingdom}

\date{\today}

\begin{abstract}
Thermal fluctuations play a central role in fluid dynamics at mesoscopic scales and must be incorporated into numerical schemes in a manner consistent with statistical mechanics. In this work, we develop a fluctuating lattice Boltzmann formulation based on an orthogonal central-moments-based representation. Stochastic forcing is introduced directly in the space of central moments (CMs) and consistently paired with mode-dependent relaxation, yielding a discrete kinetic model that satisfies the fluctuation--dissipation theorem exactly at the lattice level. Owing to the orthogonality of the basis, the equilibrium covariance matrix of the central moments is diagonal, and each non-conserved mode can be interpreted as an independent discrete Ornstein–Uhlenbeck process with variance fixed by equilibrium thermodynamics.

The resulting formulation guarantees exact equipartition of kinetic energy at equilibrium, preserves Galilean invariance, and retains the enhanced numerical stability characteristic of CMs-based collision operators. Explicit fluctuating schemes are constructed for the D2Q9 and D3Q27 lattices. The extension to reduced-velocity discretisation is discussed too. A comprehensive set of numerical tests verifies correct thermalisation, isotropy of equilibrium statistics, and the expected scaling of velocity fluctuations with thermal energy, density, and relaxation time. In contrast to fluctuating BGK formulations, the present method remains stable and well posed in the over-relaxation regime, including in the immediate vicinity of the stability limit. 

These results demonstrate that CMs-based lattice Boltzmann methods provide a natural and robust framework for fluctuating hydrodynamics, in which dissipation, noise, and kinetic mode structure are consistently aligned at the discrete level.

\end{abstract}

\maketitle

\nomenclature[A]{LBM}{lattice Boltzmann method}
\nomenclature[A]{FLBM}{fluctuating lattice Boltzmann method}
\nomenclature[A]{CMs}{central moments}
\nomenclature[A]{MRT}{multiple-relaxation-time}
\nomenclature[A]{BGK}{Bhatnagar-Gross-Krook}
\nomenclature[A]{FDT}{fluctuation--dissipation theorem}

\nomenclature[C]{$\star$}{post-collision quantity}
\nomenclature[C]{$eq$}{equilibrium quantity}
\nomenclature[D]{$i$, $j$, $h$, $v$}{indices spanning the lattice directions}
\nomenclature[D]{$\alpha$, $\beta$, $\gamma$, $\delta$}{subscripts spanning the Cartesian spatial directions}

\nomenclature[E]{$t$}{time}
\nomenclature[E]{$\mathbf{u}$}{velocity vector}
\nomenclature[E]{$\rho$}{density}
\nomenclature[E]{$\nu$}{kinematic viscosity}
\nomenclature[E]{$f_i$}{particle distribution in the direction $\bm{c}_i$}
\nomenclature[E]{$\bm{c}_i$}{lattice directions}
\nomenclature[E]{$\bar{\bm{c}}_i$}{lattice directions shifted by the local fluid velocity}
\nomenclature[E]{$w_i$}{weighting factor in the direction $\bm{c}_i$}
\nomenclature[E]{$c_s$}{lattice sound speed equal to $1/\sqrt{3}$}
\nomenclature[E]{$\tau$}{relaxation time}
\nomenclature[E]{$\omega$}{relaxation frequency}
\nomenclature[E]{$\boldsymbol \Lambda$}{relaxation matrix}
\nomenclature[E]{$\mathbf{I}$}{unit tensor}
\nomenclature[E]{$\mathbf{T}$}{transformation matrix: populations to central moments}
\nomenclature[E]{$\mathbf{M}$}{transformation matrix: populations to raw moments}
\nomenclature[E]{$\mathbf{N}$}{transformation matrix: central moments to raw moments}
\nomenclature[E]{$\mathcal{H}_{i,\alpha}^{(n)}$}{Hermite polynomial of order $n$ associated to the lattice direction $\bm{c}_i$ and the spatial component $\alpha$}
\nomenclature[E]{$\boldsymbol{\Lambda}$}{diagonal relaxation matrix in central-moment space}
\nomenclature[E]{$\tilde{\mathbf{I}}$}{diagonal selector matrix for stochastic forcing}
\nomenclature[E]{$\boldsymbol{\eta}$}{independent standard normal random variable}
\nomenclature[E]{$\boldsymbol{\phi}$}{noise amplitude}
\nomenclature[E]{$k_B\, T$}{thermal energy per lattice site}
\nomenclature[E]{$(C^{eq})_{jj}$}{equilibrium variance of the $j$-th mode}
\nomenclature[E]{$\mathbf{k}$}{central moments}
\nomenclature[E]{$\mathbf{r}$}{raw moments}
\nomenclature[E]{$b_v$}{normalisation factor for the $v$-th mode/moment}
\printnomenclature

\section{Introduction}
Thermal fluctuations are an intrinsic feature of fluid motion at mesoscopic and microscopic scales, where stochastic effects coexist with deterministic hydrodynamic transport~\cite{polimeno2025thermodynamic}. In such regimes, fluctuations influence equilibrium properties, transport coefficients, and spatio-temporal correlations, and must be accounted for to obtain physically consistent descriptions. Numerical methods aiming to resolve these phenomena are therefore required to respect both hydrodynamic conservation laws and the fluctuation--dissipation theorem (FDT)~\cite{donev2010accuracy}.\\
\indent The lattice Boltzmann method (LBM), derived from a discrete kinetic description, provides a natural framework for incorporating thermal fluctuations at the mesoscopic level~\cite{Adhikari_2005}. In LBM, fluid motion is represented by collections of particle distribution functions (also know as populations) that stream and collide on a fixed Cartesian lattice~\cite{KRUGER_Book_2017}. By augmenting the kinetic dynamics with stochastic forcing, fluctuating lattice Boltzmann methods (FLBMs) are able to reproduce equilibrium velocity and stress fluctuations and recover fluctuating hydrodynamics in the macroscopic limit~\cite{PhysRevE.76.036704}. Over the past two decades, a variety of fluctuating LB formulations have been proposed, most of them based on single--relaxation--time or raw--moment multi--relaxation--time collision operators~\cite{PhysRevE.91.023313}. While successful in many applications, these approaches often suffer from limited robustness and degraded accuracy in regimes characterised by low viscosity, strong non-equilibrium effects, or stringent stability requirements.\\
\indent Indeed, the choice of collision operator plays a central role in this context. Single--relaxation--time (also know as BGK, that is the acronym for Bhatnagar--Gross--Krook) models~\cite{BGK} offer conceptual simplicity but provide limited control over non-hydrodynamic modes, whose uncontrolled dynamics may contaminate fluctuation spectra and compromise numerical stability. Multi--relaxation--time (MRT) formulations~\cite{d2002multiple} introduce additional flexibility through mode-dependent relaxation, yet when expressed in terms of raw moments they may exhibit spurious couplings between hydrodynamic and higher-order modes, particularly at finite velocities. These limitations are amplified in fluctuating settings, where improper mode coupling or inconsistent noise allocation readily leads to violations of equipartition, distorted fluctuation spectra, and unphysical correlations~\cite{PhysRevE.94.033302}.\\
\indent A significant advance in this direction has been achieved very recently by Lauricella \textit{et al.}~\cite{10.1063/5.0288232}, who introduced a regularised fluctuating lattice Boltzmann method based on a full Hermite expansion on the D3Q27 lattice. By reconstructing both equilibrium and non-equilibrium populations on an orthogonal Hermite basis and carefully controlling the relaxation of hydrodynamic and non-hydrodynamic (ghost) modes, this approach restores thermodynamic consistency and significantly improves numerical stability across a broad range of relaxation parameters. These results clearly demonstrate that accurate fluctuating hydrodynamics at the lattice level crucially relies on orthogonal mode decompositions and a clean separation between physical and non-physical degrees of freedom.\\
\indent This insight has been further sharpened in a subsequent work by the same authors~\cite{lauricella2026ghost}, who proposed a ghost-mode filtered fluctuating lattice Boltzmann formulation. In this approach, ghost modes are prevented from carrying deterministic information altogether and are reduced to purely stochastic carriers of thermal noise, while hydrodynamic modes retain their physical dynamics. Remarkably, this minimalistic treatment yields fluctuation statistics comparable to fully regularised high-order schemes, underscoring that diagonal equilibrium covariance and strict mode decoupling are not merely advantageous, but essential for physically consistent fluctuating lattice Boltzmann formulations.\\
\indent Central-moments-based lattice Boltzmann methods (CM-LBMs) provide a natural and robust framework in which these requirements can be satisfied intrinsically~\cite{geier2006cascaded, asinari2008generalized}. By formulating the collision process in a reference frame moving with the local fluid velocity, such schemes improve Galilean invariance, suppress spurious velocity-dependent couplings, and significantly enhance numerical stability~\cite{SHAN_PRE_100_2019, COREIXAS_RSTA_378_2020}. These advantages have been demonstrated in a wide range of demanding applications, including turbulent, multiphase, and strongly non-equilibrium flows~\cite{FEI_PRE_97_2018, saito2018color}. Despite their favourable properties, the systematic construction of fluctuating lattice Boltzmann schemes within a central-moments framework has received comparatively limited attention.\\
\indent Actually, incorporating thermal fluctuations into a CMs-based formulation poses specific challenges. Stochastic forcing must be introduced directly in moment space in a manner consistent with mode-dependent relaxation, while ensuring exact equilibrium statistics and strict compliance with the FDT. At the same time, the defining properties of central-moments schemes, including numerical robustness, must be preserved.\\
\indent The aim of this work is to develop a fluctuating lattice Boltzmann formulation based on central moments that is both systematic and physically consistent. Stochastic collision operators are derived directly in space of central moments (CMs), with noise terms assigned mode by mode according to equilibrium statistical mechanics. The resulting formulation guarantees exact equipartition at equilibrium, recovers the correct fluctuating hydrodynamic behaviour in the macroscopic limit, and remains fully compatible with standard CMs bases and lattice discretisations.\\
\indent The proposed approach retains the salient features of conventional CM-LBMs and is readily applicable to large-scale simulations~\cite{ZHOU2025109686}. Numerical tests are presented to assess equilibrium fluctuations, moment-wise energy partitioning, and robustness over a broad range of relaxation parameters. The results demonstrate that central-moments-based fluctuating lattice Boltzmann methods (CM-FLBM) provide accurate and stable descriptions of thermally fluctuating flows, including regimes in which traditional fluctuating BGK formulations become unreliable.\\
\indent One of the key results of this work is that orthogonal central moments are not merely advantageous, but required to realise fluctuating hydrodynamics with diagonal equilibrium covariance and statistically independent stochastic modes at the lattice level. While correlated stochastic forcing could in principle restore the fluctuation--dissipation balance in non-orthogonal bases, the resulting noise covariance would depend explicitly on the local velocity, undermining Galilean invariance and obscuring the physical interpretation of individual modes. Unlike ghost-mode filtered formulations, which enforce statistical consistency by suppressing deterministic ghost-mode dynamics, the present approach retains a full central-moment kinetic structure and achieves diagonal equilibrium covariance intrinsically, without discarding deterministic degrees of freedom.\\
\indent The specific contributions of this work are as follows.\\
(i) A general and systematic procedure for incorporating thermal fluctuations into central-moments-based lattice Boltzmann schemes is presented, in which stochastic forcing is introduced directly into the post-collision state in terms of central moments and consistently paired with mode-dependent relaxation.\\
(ii) The formulation ensures exact equilibrium equipartition and compliance with the fluctuation--dissipation theorem for isothermal, weakly compressible flows, while preserving the defining properties of central-moments collision operators.\\
(iii) Explicit fluctuating collision operators are derived for the D2Q9 and D3Q27 lattices, providing concrete and reproducible implementations in two and three spatial dimensions. The extension to reduced-velocity discretisation is also addressed.\\
(iv) The schemes are assessed through numerical tests focusing on equilibrium velocity fluctuations, moment-wise energy partitioning, and robustness over a broad range of relaxation parameters. Extensions to multiphase, multicomponent, or fully thermal formulations are not considered here and are left for future work.

\section{Fluctuating lattice Boltzmann formulation}
\label{sec:lbe}
The construction of a fluctuating lattice Boltzmann method  requires more than the formal addition of random noise to the standard collision operator. Thermal fluctuations must simultaneously (i) preserve exact conservation of mass and momentum, (ii) satisfy a discrete fluctuation--dissipation balance at the lattice level, and (iii) remain invariant under Galilean transformations. These requirements cannot, in general, be met directly in population space, where kinetic degrees of freedom are intrinsically coupled. The strategy adopted here is therefore to reformulate the collision step in a suitably chosen moment space, where conserved, hydrodynamic, and purely kinetic modes can be clearly separated and treated independently.

\subsection{BGK fluctuating LBM}
In absence of body forces, the discrete kinetic equation can be written as~\cite{KRUGER_Book_2017}
\begin{equation}\label{eq:lbe}
f_i(\bm{x}+\bm{c}_i \Delta t,t+\Delta t) = f_i(\bm{x},t) + \Omega_i(\bm{x},t) + \zeta_i(\bm{x},t) ,
\end{equation}
where $f_i(\bm{x},t)$ denotes the particle distribution function associated with discrete velocity $\bm{c}_i$, with $i=0,\ldots,p-1$, at spatial location $\bm{x}$ and time $t$. The process is usually split into two sub-steps which are known as collision
\begin{equation}
f_i^{\star}(\bm{x},t) = f_i(\bm{x},t) + \Omega_i(\bm{x},t) + \zeta_i(\bm{x},t) ,
\end{equation}
and streaming
\begin{equation}\label{streaming}
f_i(\bm{x}+\bm{c}_i \Delta t,t+\Delta t) = f_i^{\star}(\bm{x},t),
\end{equation}
respectively, with $\Delta t=1$. The superscript $\star$ denotes post-collision quantities. To lighten the notation, let us drop the dependence on space and time in the rest of this section. 

In its simplest form, the deterministic collision operator is expressed as a relaxation towards a local equilibrium~\cite{BGK}, \textit{i.e.}
\begin{equation}
\Omega_i
=
\omega\left(f_i^{eq}-f_i\right),
\end{equation}
where the relaxation frequency $\omega$ is related to the kinematic viscosity $\nu$ through
\begin{equation}
\omega
=
\left(\frac{\nu}{c_s^2}+\frac{1}{2}\right)^{-1},
\end{equation}
and $c_s$ is the lattice sound speed. Macroscopic density and velocity fields are recovered as
\begin{equation} \label{eq:macro}
\rho=\sum_i f_i,
\qquad
\rho\bm{u}=\sum_i \bm{c}_i f_i,
\end{equation}
respectively. Thermal fluctuations are usually introduced through a stochastic term, namely $\zeta_i$. Rather than injecting noise directly at the population level, fluctuations are generated in kinetic moment space and subsequently projected back onto the populations. This approach allows fluctuations to be applied selectively to non-conserved degrees of freedom while preserving the exact conservation laws. The stochastic contribution to the populations is written as
\begin{equation}
\zeta_i
=
w_i
\sum_v
b_v^{-1}
e_{vi}\,
\phi_v\,
\eta_v ,
\label{eq:Xi_pop}
\end{equation}
where $\mathbf{e}_v=[e_{v0},\ldots,e_{v\,p-1}]$ denotes the basis vector associated with the $v$-th kinetic mode, $b_v=\sum_i w_i e_{vi}^2$ is its normalisation factor, $w_i$ is the lattice weight, and $\eta_v$ is an independent standard normal variable. The fluctuation amplitude $\phi_v$ is determined by the discrete fluctuation--dissipation theorem. Let us underline that the index $k$ in Eq.~(\ref{eq:Xi_pop}) does not span the whole set of degrees of freedom (also known as modes or moments), but only the non-conserved ones~\cite{10.1063/5.0288232}. 

Interpreting the stochastic collision operator as a Monte Carlo process satisfying detailed balance, the fluctuation amplitude takes the form
\begin{equation}
\phi_v
=
\sqrt{
\frac{
\rho\,k_B T\,\Lambda_v(2-\Lambda_v)\,b_v
}{
c_s^2
}
},
\label{eq:phi_general}
\end{equation}
where $\Lambda_v$ is the relaxation rate of the $v$-th mode. This choice ensures exact balance between dissipation and fluctuations at the lattice level and recovers the Landau--Lifshitz fluctuating Navier--Stokes equations in the hydrodynamic limit.

\subsection{Central-moments-based fluctuating LBM: Two-dimensional formulation}

We consider a two-dimensional Cartesian system $(x,y)$ and the D2Q9 lattice ($p=9$) with discrete velocities $\bm{c}_i=[c_{i,x},c_{i,y}]$,
\begin{eqnarray}
c_{i,x} &=& [0, 1, 0, -1, 0, 1, -1, -1, 1], \\
c_{i,y} &=& [0, 0, 1, 0, -1, 1, 1, -1, -1]. \nonumber
\end{eqnarray}
The flow velocity vector is $\bm{u} = [u_x, u_y]$. The equilibrium distribution $f_i^{eq}$ is expressed via a Hermite expansion~\cite{PhysRevE.96.033306} 
\begin{equation} \label{eq:feq}
f_i^{eq} = w_i \sum_{n=0}^N \frac{1}{n! c_s^{2n}} \mathcal{H}_{i, \boldsymbol{\alpha}}^{(n)} a_{0, \boldsymbol{\alpha}}^{(n)}, 
\end{equation} 
where the corresponding lattice weights are
\begin{equation}
w_0=\frac{4}{9},
\quad
w_{1\text{--}4}=\frac{1}{9},
\quad
w_{5\text{--}8}=\frac{1}{36},
\end{equation}
and the definition of the Hermite tensor up to the order $N=4$ is 
\begin{eqnarray} \mathcal{H}_i^{(0)} &=& 1, \\ \nonumber 
\mathcal{H}_{i, \alpha}^{(1)} &=& c_{i, \alpha}, \\ \nonumber 
\mathcal{H}_{i, \alpha \beta}^{(2)} &=& c_{i, \alpha \beta} - c_s^2 \delta_{\alpha \beta}, \\ \nonumber 
\mathcal{H}_{i, \alpha \beta \gamma}^{(3)} &=& c_{i, \alpha \beta \gamma} - c_s^2 \left( c_{i, \alpha} \delta_{\beta \gamma} + c_{i, \beta} \delta_{\alpha \gamma} + c_{i, \gamma} \delta_{\alpha \beta} \right), \\ \nonumber 
\mathcal{H}^{(4)}_{i,\alpha\beta\gamma\delta} &=& c_{i,\alpha} c_{i,\beta} c_{i,\gamma} c_{i,\delta} - c_s^2 \Big( c_{i,\alpha}c_{i,\beta}\delta_{\gamma\delta} + c_{i,\alpha}c_{i,\gamma}\delta_{\beta\delta}\\ \nonumber &+& c_{i,\alpha}c_{i,\delta}\delta_{\beta\gamma} + c_{i,\beta}c_{i,\gamma}\delta_{\alpha\delta} \\ \nonumber &+& c_{i,\beta}c_{i,\delta}\delta_{\alpha\gamma} + c_{i,\gamma}c_{i,\delta}\delta_{\alpha\beta} \Big) \\ \nonumber &+& c_s^4 \Big( \delta_{\alpha\beta}\delta_{\gamma\delta} + \delta_{\alpha\gamma}\delta_{\beta\delta} + \delta_{\alpha\delta}\delta_{\beta\gamma} \Big), \end{eqnarray} 
with $c_s=1/\sqrt{3}$, $c_{i, \alpha \beta} = c_{i, \alpha} c_{i, \beta}$, $c_{i, \alpha \beta \gamma} = c_{i, \alpha} c_{i, \beta} c_{i, \gamma}$ and $c_{i, \alpha \beta \gamma \delta} = c_{i, \alpha} c_{i, \beta} c_{i, \gamma} c_{i, \delta}$, $\delta$ is the Kronecker delta, and $\left( \alpha, \beta, \gamma, \delta \right) \in \{x,y\}^4$. The corresponding Hermite coefficients are 
\begin{eqnarray} 
a_0^{(0)} &=& \rho, \\ \nonumber 
a_{0, \alpha}^{(1)} &=& u_{\alpha}a_0^{(0)} = u_{\alpha} \rho, \\ \nonumber 
a_{0, \alpha \beta}^{(2)} &=& u_{\beta} a_{0, \alpha}^{(1)} = u_{\beta} u_{\alpha} \rho, \\ \nonumber
a_{0, \alpha \beta \gamma}^{(3)} &=& u_{\gamma} a_{0, \alpha \beta}^{(2)}= u_{\gamma} u_{\beta} u_{\alpha} \rho, \\ \nonumber 
a_{0, \alpha \beta \gamma \delta }^{(4)} &=& u_{\delta} a_{0, \alpha \beta \gamma}^{(3)} = u_{\delta} u_{\gamma} u_{\beta} u_{\alpha} \rho. 
\end{eqnarray} 
Equilibrium distributions can be compactly rewritten as follows:
\begin{eqnarray}
    f_i^{eq} = w_i &\rho& \Big[1 + \frac{\bm{c}_i \cdot \bm{u}}{c_s^2} + \frac{1}{2c_s^4} \mathcal{H}_i^{(2)}:\bm{u}\bm{u} + \\
    & & \frac{1}{2c_s^6}\left( \mathcal{H}_{i, xxy}^{(3)} u_x^2u_y + \mathcal{H}_{i, xyy}^{(3)} u_x u_y^2 \right) + \nonumber \\
    & & \mathcal{H}^{(4)}_{i,xxyy} u_x^2 u_y^2  \Big]. \nonumber 
\end{eqnarray}
Note that $N=4$ is the maximum order attainable in the D2Q9 lattice. If one sets $N=2$, the classical second-order truncated expression is immediately recovered.\\
\indent To separate conserved, non-conserved, and purely kinetic degrees of freedom, particle populations are mapped onto moments. While standard multiple-relaxation-time lattice Boltzmann formulations are commonly expressed in terms of raw moments, it has been shown that CMs-based formulations provide superior numerical robustness and improved Galilean invariance~\cite{FLD:FLD4208}. Central moments of order $(l,m)$ can be generally defined as
\begin{equation}
k_{\alpha^l\beta^m}
=
\sum_i
f_i\,
\bar{c}_{i,\alpha}^l
\bar{c}_{i,\beta}^m,
\qquad
\bar{c}_{i,\alpha}=c_{i,\alpha}-u_\alpha .
\end{equation}
The combination of $\bar{c}_{i,\alpha}^l \bar{c}_{i,\beta}^m$ are not arbitrary, but they aim at building a basis (that is \textit{just} a square matrix) where each row corresponds to a specific kinetic moment/mode with a clear physical interpretation. Before going any further, it is more convenient to rearrange some relevant quantities by gathering them into vectors. To this end, let us introduce the vectors $\mathbf{k}=[k_0, \ldots, k_{p-1}]$ and 
$\mathbf{f}=[f_0, \ldots, f_{p-1}]$ which collect central moments and populations, respectively. The mapping between the two can be formulated as
\begin{equation}
\mathbf{k} =\mathbf{T} \, \mathbf{f}.
\end{equation}
Notably, this basis can be rewritten as $\mathbf{T}=\mathbf{N}\mathbf{M}$, where $\mathbf{M}$ is fixed, constant, and  transform populations to raw moments, while $\mathbf{N}=\mathbf{M}^{-1} \mathbf{T}$ contains local-velocity-dependent terms and performs the shift from raw to central moments.

The collision step is formulated in the CMs space as
\begin{equation}
\mathbf{k}^\star = (\mathbf{I}-\boldsymbol{\Lambda})\, \mathbf{k} + \boldsymbol{\Lambda}\, \mathbf{k}^{eq} + \tilde{\mathbf{I}} \,\boldsymbol{\phi}\, \boldsymbol{\eta},
\label{eq:cm_collision}
\end{equation}
where $\displaystyle \mathbf{k}^{eq} = \mathbf{T} \, \mathbf{f}^{eq}$ and $\mathbf{k}^\star$ collect the equilibrium and post-collision central moments, respectively, $\mathbf{I}$ is the identity matrix, $\boldsymbol{\Lambda}$ is the diagonal relaxation matrix, and $\tilde{\mathbf{I}}$ is a diagonal matrix allowing us to apply noise to some selected modes. Fluctuation amplitudes in Eq.~(\ref{eq:phi_general}) and standard normal variables are collected by $\boldsymbol{\phi}$ and $\boldsymbol{\eta}$, respectively. For each non-conserved mode, Eq.~(\ref{eq:cm_collision}) defines a discrete-time Ornstein--Uhlenbeck process in the CMs space. Stationarity requires the variance of this process to match the equilibrium covariance $(C^{eq})_j$, which uniquely determines the noise amplitude $\phi_j$.\\
\indent For the FDT to be enforced mode by mode, the equilibrium covariance matrix of the kinetic modes must be diagonal, \textit{i.e.}
\begin{equation}
\langle \delta k_j \, \delta k_h \rangle = \left(C^{eq}\right)_j \, \delta_{jh},
\end{equation}
where $\delta k_j = k_j - \langle k_j \rangle$ denotes the fluctuation of the $j$-th central moment around its equilibrium mean (and indices $j$ and $h$ here label kinetic modes).\\
\indent Several lattice Boltzmann formulations have been proposed based on non-orthogonal moment bases, motivated by their algebraic simplicity and ease of implementation~\cite{10.1063/1.5087266}. Therefore, let us first employ a commonly used non-orthogonal basis defined as~\cite{derosis2016epl_d2q9}
\begin{eqnarray}
M_{0 i} &=& 1, \\
M_{1 i} &=& c_{i,x}, \nonumber \\
M_{2 i} &=& c_{i,y}, \nonumber \\
M_{3 i} &=& c_{i,x}^2+c_{i,y}^2, \nonumber \\
M_{4 i} &=& c_{i,x}^2-c_{i,y}^2,  \nonumber\\
M_{5 i} &=& c_{i,x}c_{i,y}, \nonumber \\
M_{6 i} &=& c_{i,x}^2c_{i,y}, \nonumber \\
M_{7 i} &=& c_{i,x}c_{i,y}^2, \nonumber \\
M_{8 i} &=& c_{i,x}^2c_{i,y}^2. \nonumber
\end{eqnarray}
While such choice is adequate for deterministic simulations, it poses fundamental difficulties in the presence of thermal fluctuations. In non-orthogonal bases, the equilibrium covariance matrix of kinetic modes is generally non-diagonal, implying statistical correlations between different moments at equilibrium. As a consequence, stochastic forcing cannot be applied independently to each mode without violating the discrete FDT. Although correlated noise could in principle be constructed to compensate for this coupling, such procedures obscure the physical interpretation of the modes and complicate both analysis and implementation. By contrast, an orthogonal basis guarantees diagonal equilibrium covariance, enabling independent relaxation and stochastic forcing of each kinetic mode. Orthogonality is therefore not a matter of numerical convenience, but a necessary structural condition for enforcing fluctuation–dissipation balance in a transparent and physically consistent manner. Orthogonality of the basis is therefore required to guarantee statistical independence of the modes and to allow uncorrelated stochastic forcing.\\
\indent Accordingly, we propose to adopt the following orthogonal basis
\begin{eqnarray}
M_{0 i} &=& 1,  \\
M_{1 i} &=& c_{i,x}, \nonumber \\
M_{2 i} &=& c_{i,y},  \nonumber\\
M_{3 i} &=& c_{i,x}^2+c_{i,y}^2-2c_s^2, \nonumber \\
M_{4 i} &=& c_{i,x}^2-c_{i,y}^2,  \nonumber\\
M_{5 i} &=& c_{i,x}c_{i,y},  \nonumber\\
M_{6 i} &=& (c_{i,x}^2-c_s^2)c_{i,y},  \nonumber\\
M_{7 i} &=& c_{i,x}(c_{i,y}^2-c_s^2), \nonumber \\
M_{8 i} &=& (c_{i,x}^2-c_s^2)(c_{i,y}^2-c_s^2).\nonumber
\end{eqnarray}

The shifted transformation matrix $\mathbf{T}$ is obtained by replacing $c_{i,\alpha}$ by $\bar{c}_{i,\alpha}$. In this basis, the equilibrium covariance matrix does not show any off-diagonal term and, conveniently, all equilibrium central moments vanish except for 
\begin{equation}
    k_0^{eq}=\rho.
\end{equation}
The corresponding normalisation factors are
\begin{eqnarray}
&b_0&=1,\quad
b_{1-2}=\frac{1}{3},\quad
b_{3-4}=\frac{4}{9},\quad  \\
&b_5&=\frac{1}{9},\quad
b_{6-7}=\frac{2}{27},\quad
b_8=\frac{4}{81}.\nonumber
\end{eqnarray}

With this basis, the collision step in Eq.~(\ref{eq:cm_collision}), equipped by $\boldsymbol{\Lambda} = \mathrm{diag}[1, 1, 1, 1, \omega, \omega, 1, 1, 1]$ and $\tilde{\mathbf{I}} = \mathrm{diag}[0, 0, 0, 1, 1, 1, 1, 1, 1]$, results in
\begin{eqnarray} \label{eq:post_coll_cms}
    k_0^{\star} &=& \rho,\\
    k_1^{\star} &=& 0, \nonumber \\
    k_2^{\star} &=& 0, \nonumber \\
    k_3^{\star} &=& 2 \sqrt{\rho \, k_B \, T c_s^2}\, \eta_3,\nonumber \\
    k_4^{\star} &=& (1-\omega)k_4 + 2 \sqrt{\rho \, k_B \, T \,  c_s^2 \, \omega (2-\omega)} \, \eta_4, \nonumber \\
    k_5^{\star} &=& (1-\omega)k_5 + \sqrt{\rho \, k_B \, T \,c_s^2 \, \omega (2-\omega)} \, \eta_5, \nonumber \\
    k_6^{\star} &=& \sqrt{2 \rho \, k_B \, T}\, c_s^2 \,\eta_6, \nonumber \\
    k_7^{\star} &=& \sqrt{2 \rho \, k_B \, T} \,c_s^2 \, \eta_7, \nonumber \\
    k_8^{\star} &=& 2 \sqrt{\rho \, k_B \, T}\, c_s^3 \,\eta_8. \nonumber 
\end{eqnarray}
The numerical prefactors in the above-listed equations follow directly from the modal norms $b_v$ (see Eq.~(\ref{eq:Xi_pop})) associated with our orthogonal basis. The only relevant pre-collision CMs are
\begin{eqnarray} \label{eq:pre_coll_cms}
    k_4 &=& r_4 -\rho(u_x^2-u_y^2), \\
    k_5 &=& r_5 -\rho u_x u_y, \nonumber
\end{eqnarray}
where the two pre-collision raw moments involved in the computations are
\begin{eqnarray} \label{eq:pre_coll_rms}
    r_4 &=& f_1-f_2+f_3-f_4, \\
    r_5 &=& f_5-f_6+f_7-f_8. \nonumber
\end{eqnarray}
Post-collision populations can be computed by applying the back transformation
\begin{equation}
\mathbf{f}^{\star} = \mathbf{T}^{-1} \mathbf{k}^{\star}.
\end{equation}
However, it is more convenient from a pure computational viewpoint to first compute the post-collision raw moments
\begin{equation}\label{eq:post_coll_rms_compact}
\mathbf{r}^{\star} = \mathbf{N}^{-1} \mathbf{k}^{\star},
\end{equation}
which are
\begin{eqnarray}\label{eq:post_coll_rms}
    r_0^{\star} &=& k_0^{\star}, \\
    r_1^{\star} &=& \rho  u_x, \nonumber \\
    r_2^{\star} &=& \rho  u_y, \nonumber \\
    r_3^{\star} &=& k_3^{\star} + \rho(u_x^2+u_y^2), \nonumber \\
    r_4^{\star} &=& k_4^{\star}+\rho (u_x^2-u_y^2), \nonumber \\
    r_5^{\star} &=& k_5^{\star}+\rho u_x u_y, \nonumber \\
    r_6^{\star} &=& k_6^{\star}+2u_x k_5^{\star} + \frac{1}{2}u_y \left( k_3^{\star}+k_4^{\star} \right) + \rho u_x^2 u_y, \nonumber \\
    r_7^{\star} &=& k_7^{\star}+2u_y k_5^{\star} + \frac{1}{2} u_x \left( k_3^{\star}-k_4^{\star} \right) + \rho u_x u_y^2, \nonumber \\
    r_8^{\star} &=& k_8^{\star}+2 \left( u_x k_7^{\star} +u_y k_6^{\star}  \right)+ 4 u_x u_y k_5^{\star} \nonumber \\
    &-&\frac{1}{2}(u_x^2-u_y^2) k_4^{\star} + \frac{1}{2}(u_x^2+u_y^2) k_3^{\star}  + \rho u_x^2 u_y^2,\nonumber 
\end{eqnarray}
and then transform these into post-collision populations by 
\begin{equation}\label{eq:post_coll_pops_compact}
\mathbf{f}^{\star} = \mathbf{M}^{-1} \mathbf{r}^{\star},
\end{equation}
resulting in
\begin{eqnarray}\label{eq:post_coll_pops}
f_0^{\star} &=& \frac{4 r_0^{\star}}{9}  - \frac{2 r_3^{\star}}{3}  + r_8^{\star}, \\
f_1^{\star} &=& \frac{r_0^{\star}}{9}  + \frac{r_1^{\star}}{3}  + \frac{r_3^{\star}}{12} 
+ \frac{r_4^{\star}}{4}  - \frac{r_7^{\star} + r_8^{\star}}{2} , \nonumber \\
f_2^{\star} &=& \frac{r_0^{\star}}{9}  + \frac{r_2^{\star}}{3}  + \frac{r_3^{\star}}{12} 
- \frac{r_4^{\star}}{4}  - \frac{r_6^{\star} + r_8^{\star}}{2} , \nonumber \\
f_3^{\star} &=& \frac{r_0^{\star}}{9}  - \frac{r_1^{\star}}{3}  + \frac{r_3^{\star}}{12} 
+ \frac{r_4^{\star}}{4}  + \frac{r_7^{\star} - r_8^{\star}}{2} , \nonumber \\
f_4^{\star} &=& \frac{r_0^{\star}}{9}  - \frac{r_2^{\star}}{3} + \frac{r_3^{\star}}{12} 
- \frac{r_4^{\star}}{4}  + \frac{r_6^{\star} - r_8^{\star}}{2}, \nonumber \\
f_5^{\star} &=& \frac{r_0^{\star}}{36}  + \frac{r_1^{\star} + r_2^{\star} +  r_3^{\star}}{12}
+ \frac{r_5^{\star} + r_6^{\star} + r_7^{\star} + r_8^{\star}}{4}, \nonumber \\
f_6^{\star} &=& \frac{r_0^{\star}}{36}  + \frac{-r_1^{\star} + r_2^{\star} + r_3^{\star}}{12}
+ \frac{-r_5^{\star} +  r_6^{\star} - r_7^{\star} + r_8^{\star}}{4}, \nonumber \\
f_7^{\star} &=& \frac{r_0^{\star}}{36}  + \frac{-r_1^{\star} - r_2^{\star} + r_3^{\star}}{12}
+ \frac{r_5^{\star} -  r_6^{\star} - r_7^{\star} + r_8^{\star}}{4}, \nonumber \\
f_8^{\star} &=& \frac{r_0^{\star}}{36} + \frac{r_1^{\star} - r_2^{\star} + r_3^{\star}}{12}
+ \frac{-r_5^{\star} -  r_6^{\star} + r_7^{\star} + r_8^{\star}}{4},\nonumber 
\end{eqnarray}
which are eventually streamed as usual by Eq.~(\ref{streaming}).

\subsubsection*{Structural properties of the central-moment equilibrium}

A defining feature of central-moments-based lattice Boltzmann methods (CM-LBMs) is that the transformation matrix $\mathbf{T}$ can be interpreted as a discrete Hermite basis centred on the local fluid velocity. Projection of the particle populations onto this basis yields a natural hierarchy of central moments with a clear physical interpretation: $k_0$ represents the density, $k_{1,2}$ correspond to the (vanishing) central momenta, $k_{3,4,5}$ span the second-order stress sector associated with viscous transport, and $k_{6,7,8}$ represent higher-order kinetic (ghost) modes.

When the equilibrium distribution is constructed consistently up to fourth order in the Hermite expansion, the equilibrium manifold in central-moment space acquires a particularly simple and physically transparent structure. In the present case, all equilibrium central moments vanish identically except for the density,
\begin{equation}
k_0^{eq} = \rho .
\end{equation}
This reflects the fact that, in the local rest frame, thermodynamic equilibrium does not populate higher-order central moments. Consequently, the deterministic equilibrium structure is entirely confined to the conserved subspace, while all non-conserved modes fluctuate about zero.

By contrast, if a second-order (classical) equilibrium distribution is employed, this property is lost. Truncation of the Hermite expansion induces non-zero, velocity-dependent equilibrium values in higher-order central moments. In the present D2Q9 basis, the equilibrium central-moment vector takes the form
\begin{equation}
\mathbf{k}^{eq}=
\begin{pmatrix}
\rho\\
0\\
0\\
0\\
0\\
0\\
-\,\rho\,u_x^{2}u_y\\
-\,\rho\,u_x u_y^{2}\\
3\,\rho\,u_x^{2}u_y^{2}
\end{pmatrix},
\end{equation}
so that ghost modes acquire deterministic equilibrium offsets. The equilibrium manifold therefore extends into the kinetic subspace, and higher-order modes are no longer centred about zero even at equilibrium.

It is important to emphasise that the appearance of non-zero equilibrium values in higher-order central moments under a second-order equilibrium is not a numerical artefact, nor a consequence of discretisation errors, finite resolution, or implementation details. Rather, it is a structural consequence of truncating the Hermite expansion of the equilibrium distribution. Such truncation inevitably projects velocity-dependent contributions onto higher-order central moments, thereby contaminating the kinetic subspace with deterministic equilibrium structure. This behaviour persists independently of lattice resolution, time step, or numerical precision. By contrast, when the equilibrium distribution is constructed up to the maximum Hermite order supported by the lattice, all non-conserved central moments vanish identically at equilibrium, yielding a clean separation between deterministic equilibrium structure and stochastic fluctuations.

This distinction has direct implications for fluctuating lattice Boltzmann formulations. When higher-order central moments possess non-zero equilibrium means, the separation between deterministic equilibrium structure and stochastic fluctuations becomes blurred, complicating both the interpretation and the calibration of thermal noise. In particular, deterministic offsets in kinetic modes undermine the physical transparency associated with a diagonal equilibrium structure in central-moment space.

To formalise this distinction, consider the second moment of the central-moment vector $\mathbf{k}$. Introducing the outer product
\begin{equation}
\mathbf{A} = \mathbf{k}^{eq}(\mathbf{k}^{eq})^{T},
\end{equation}
which represents the deterministic contribution to $\langle \mathbf{k}\mathbf{k}^{T} \rangle$, one obtains the exact decomposition
\begin{equation}
\langle \mathbf{k}\mathbf{k}^{T} \rangle
=
\underbrace{\langle \delta\mathbf{k}\,\delta\mathbf{k}^{T} \rangle}_{\mathbf{C}^{eq}}
+
\mathbf{k}^{eq}(\mathbf{k}^{eq})^{T},
\end{equation}
where $\delta\mathbf{k}=\mathbf{k}-\mathbf{k}^{eq}$. The equilibrium covariance matrix in central-moment space is therefore defined as
\begin{equation}
\mathbf{C}^{eq} = \langle \delta\mathbf{k}\,\delta\mathbf{k}^{T} \rangle.
\end{equation}

Assuming local equilibrium population fluctuations of ideal-gas form,
\begin{equation}
\langle \delta f_i\,\delta f_j\rangle
=
\frac{\rho k_B T}{c_s^2}\, w_i\,\delta_{ij},
\end{equation}
the equilibrium covariance in central-moment space follows by linear transformation,
\begin{equation}
(C^{eq})_{v\ell}
=
\frac{\rho k_B T}{c_s^2}
\sum_i T_{v i}\, w_i\, T_{\ell i}.
\end{equation}
For the orthogonal central-moment basis adopted here, the transformation matrix $\mathbf{T}$ satisfies a weighted orthogonality relation with respect to the lattice quadrature,
\begin{equation}
\sum_i w_i\, T_{v i}\, T_{\ell i}
=
b_v\,\delta_{v\ell},
\end{equation}
where $b_v$ are the modal norms evaluated in the local rest frame. As a direct consequence, the equilibrium covariance matrix is strictly diagonal,
\begin{equation}
(C^{eq})_{v\ell}
=
\frac{\rho k_B T}{c_s^2}\, b_v\,\delta_{v\ell}.
\end{equation}

The diagonal structure of $\mathbf{C}^{eq}$ reflects the statistical independence of central moments at equilibrium and highlights the separation between hydrodynamic and kinetic subspaces. In particular, when a fourth-order equilibrium is employed, only the density mode possesses a non-zero equilibrium mean, rendering the deterministic contribution $\mathbf{A}$ rank-one and confined to the conserved sector, while all equilibrium fluctuations are fully captured by $\mathbf{C}^{eq}$. This property enables independent relaxation and stochastic forcing of non-hydrodynamic modes and provides a transparent and robust realisation of the fluctuation--dissipation balance. Such a simplification does not generally hold in raw-moment formulations, where both the deterministic contribution and the equilibrium covariance are sparse and weakly structured, irrespective of the equilibrium order. In non-orthogonal bases, enforcing the fluctuation--dissipation theorem requires correlated noise whose covariance depends explicitly on the local velocity, thereby undermining both Galilean invariance and physical interpretability.

Finally, the present fluctuating lattice Boltzmann formulation is fully consistent with the framework of Landau--Lifshitz fluctuating hydrodynamics. In the hydrodynamic limit, stochastic forcing applied to the second-order central moments produces Gaussian, white-in-time stress fluctuations whose variance is proportional to viscous dissipation, in agreement with the continuum fluctuation--dissipation theorem. Higher-order kinetic modes act as rapidly relaxing degrees of freedom that transmit fluctuations to hydrodynamic scales without introducing spurious long-range correlations. Because stochastic forcing is defined in the local rest frame through central moments, the resulting fluctuation statistics are Galilean invariant by construction. The discrete formulation therefore recovers the Landau--Lifshitz fluctuating Navier--Stokes equations asymptotically, while retaining exact mass and momentum conservation and a well-defined fluctuation--dissipation balance at the lattice level.

Summing up, an orthogonal central-moment formulation combined with a Hermite-consistent equilibrium constitutes the minimal discrete setting in which equilibrium structure, dissipation, and stochastic forcing are aligned in a statistically consistent and physically transparent manner.

\subsubsection*{Algorithm of computation}
Within the typical time step and for each lattice site, the implementation of our methodology requires the following steps:
\begin{enumerate}
    \item compute macroscopic variables by Eqs.~(\ref{eq:macro});
    \item sample the random variables $\eta_3, \ldots, \eta_8$;
    \item evaluate pre-collision central moments by Eqs.~(\ref{eq:pre_coll_cms}) using Eqs.~(\ref{eq:pre_coll_rms});
    \item perform the collision step by Eqs.~(\ref{eq:post_coll_cms});
    \item obtain post-collision raw moments from post-collision central moments by Eqs.~(\ref{eq:post_coll_rms});
    \item reconstruct post-collision populations by Eqs.~(\ref{eq:post_coll_pops});
    \item stream by Eq.~(\ref{streaming}) and advance in time.
\end{enumerate}
Notably, all steps are local and involve only scalar operations per lattice site.

\subsection{Central-moments-based fluctuating LBM: Three-dimensional formulation}
\label{sec:D3Q27}
Building on the generality of the methodology outlined above for the two-dimensional D2Q9 lattice, we can directly extend it to the three-dimensional D3Q27 discretisation. Considering a three-dimensional Cartesian system $(x,y,z)$, the lattice directions in Eq.~(\ref{eq:lbe}) now are $\bm{c}_i = [c_{i,x}, c_{i,y}, c_{i,z}]$, where
\begin{eqnarray}
    c_{i,x} &=& [0, 1, -1, 0, 0, 0, 0, 1, -1, 1, -1, 1, -1, 1, \\ 
           & & -1, 0, 0, 0, 0, 1, -1, 1, -1, 1, -1, 1, -1],\nonumber \\
    c_{i,y} &=& [0, 0, 0, 1, -1, 0, 0, 1, 1, -1, -1, 0, 0, 0, \nonumber \\
           & & 0 , 1, -1, 1, -1, 1, 1, -1, -1, 1, 1, -1, -1], \nonumber \\
    c_{i,z} &=& [0, 0, 0, 0, 0, 1, -1, 0, 0, 0, 0, 1, 1, -1, -1,  \nonumber \\
           & &  1, 1, -1, -1, 1, 1, 1, 1, -1, -1, -1, -1], \nonumber
\end{eqnarray}
with $i=0\,\ldots,p-1$, and $p=27$. Weighting factors $w_0=8/27$, $w_{1-6}=2/27$, $w_{7-18}=1/54$, $w_{19-26} = 1/216$. This lattice discretisation allows us to deploy the equilibrium populations in Eq.~(\ref{eq:feq}) up to the sixth order, that is
\begin{widetext}
\begin{equation}
\begin{aligned}
f_i^{eq} = w_i \rho \Bigg\{
& 1 + \frac{\bm{c}_i \cdot \boldsymbol{u}}{c_s^2}
+ \frac{1}{2 c_s^4} \, \mathcal{H}_i^{(2)} : \boldsymbol{u}\boldsymbol{u}
\\
& + \frac{1}{2 c_s^6} \Big(
  \mathcal{H}_{ixxy}^{(3)} u_x^2 u_y
+ \mathcal{H}_{ixxz}^{(3)} u_x^2 u_z
+ \mathcal{H}_{ixyy}^{(3)} u_x u_y^2
+ \mathcal{H}_{ixzz}^{(3)} u_x u_z^2
+ \mathcal{H}_{iyzz}^{(3)} u_y u_z^2
+ \mathcal{H}_{iyyz}^{(3)} u_y^2 u_z
+ 2 \mathcal{H}_{ixyz}^{(3)} u_x u_y u_z
\Big)
\\
& + \frac{1}{4 c_s^8} \Big[
  \mathcal{H}_{ixxyy}^{(4)} u_x^2 u_y^2
+ \mathcal{H}_{ixxzz}^{(4)} u_x^2 u_z^2
+ \mathcal{H}_{iyyzz}^{(4)} u_y^2 u_z^2
+ 2 \big(
    \mathcal{H}_{ixyzz}^{(4)} u_x u_y u_z^2
  + \mathcal{H}_{ixyyz}^{(4)} u_x u_y^2 u_z
  + \mathcal{H}_{ixxyz}^{(4)} u_x^2 u_y u_z
  \big)
\Big]
\\
& + \frac{1}{4 c_s^{10}} \Big(
  \mathcal{H}_{ixxyzz}^{(5)} u_x^2 u_y u_z^2
+ \mathcal{H}_{ixxyyz}^{(5)} u_x^2 u_y^2 u_z
+ \mathcal{H}_{ixyyzz}^{(5)} u_x u_y^2 u_z^2
\Big)
+ \frac{1}{8 c_s^{12}}
  \mathcal{H}_{ixxyyzz}^{(6)} u_x^2 u_y^2 u_z^2
\Bigg\},
\end{aligned}
\end{equation}
\end{widetext}
with the flow velocity vector being $\bm{u} = [u_x, u_y, u_z]$. Details about such formula can be found in~\cite{malaspinas2015increasingstabilityaccuracylattice, PhysRevE.96.033306}. In three dimensions, central moments of order $(l,m,q)$ are defined as
\begin{equation}
k_{\alpha^l\beta^m \gamma^q}
=
\sum_i
f_i\,
\bar{c}_{i,\alpha}^l
\bar{c}_{i,\beta}^m
\bar{c}_{i,\gamma}^q
\end{equation}

The twenty-seven post-collision central moments can be computed by Eq.~(\ref{eq:cm_collision}) providing a suitable basis~\cite{de2017nonorthogonal}. Based on the arguments about orthogonality, we are not in the position to use the $27 \times 27$ non-orthogonal matrix by De Rosis \& Coreixas~\cite{DEROSIS_PoF_31_2019}. Therefore, we employ the orthogonal one by Lauricella \textit{et al.}~\cite{10.1063/5.0288232}, that is
\begin{eqnarray}
    M_{0 i} &=& 1,  \\
    M_{1 i} &=& c_{i,x}, \nonumber \\
    M_{2 i} &=& c_{i,y},  \nonumber\\
    M_{3 i} &=& c_{i,z},  \nonumber\\
    M_{4 i} &=& c_{i,x}^2-c_s^2, \nonumber \\
    M_{5 i} &=& c_{i,y}^2-c_s^2, \nonumber \\
    M_{6 i} &=& c_{i,z}^2-c_s^2, \nonumber \\
    M_{7 i} &=& c_{i,x}c_{i,y}, \nonumber \\
    M_{8 i} &=& c_{i,x}c_{i,z}, \nonumber \\
    M_{9 i} &=& c_{i,y}c_{i,z}, \nonumber \\
    M_{10 i} &=& c_{i,x}^2c_{i,y}-c_s^2c_{i,y}, \nonumber \\
    M_{11 i} &=& c_{i,x}^2c_{i,z}-c_s^2c_{i,z}, \nonumber \\
    M_{12 i} &=& c_{i,x}c_{i,y}^2-c_s^2c_{i,x}, \nonumber \\
    M_{13 i} &=& c_{i,x}c_{i,z}^2-c_s^2c_{i,x}, \nonumber \\
    M_{14 i} &=& c_{i,y}c_{i,z}^2-c_s^2c_{i,y}, \nonumber \\
    M_{15 i} &=& c_{i,y}^2c_{i,z}-c_s^2c_{i,y}, \nonumber \\
    M_{16 i} &=& c_{i,x}c_{i,y}c_{i,z}, \nonumber \\
    M_{17 i} &=& c_{i,x}^2c_{i,y}^2 -c_s^2(c_{i,x}^2+c_{i,y}^2)+c_s^4, \nonumber \\
    M_{18 i} &=& c_{i,x}^2c_{i,z}^2 -c_s^2(c_{i,x}^2+c_{i,z}^2)+c_s^4, \nonumber \\
    M_{19 i} &=& c_{i,y}^2c_{i,z}^2 -c_s^2(c_{i,y}^2+c_{i,z}^2)+c_s^4, \nonumber \\
    M_{20 i} &=& c_{i,x}c_{i,y}c_{i,z}^2 -c_s^2c_{i,x}c_{i,y}, \nonumber \\
    M_{21 i} &=& c_{i,x}c_{i,y}^2c_{i,z} -c_s^2c_{i,x}c_{i,z}, \nonumber \\
    M_{22 i} &=& c_{i,x}^2c_{i,y}c_{i,z} -c_s^2c_{i,y}c_{i,z}, \nonumber \\
    M_{23 i} &=& c_{i,x}^2c_{i,y}c_{i,z}^2 -c_s^2(c_{i,x}^2c_{i,y}+c_{i,y}c_{i,z}^2)+c_s^4c_{i,y}, \nonumber \\
    M_{24 i} &=& c_{i,x}^2c_{i,y}^2c_{i,z} -c_s^2(c_{i,x}^2c_{i,z}+c_{i,y}^2c_{i,z})+c_s^4c_{i,z}, \nonumber \\
    M_{25 i} &=& c_{i,x}c_{i,y}^2c_{i,z}^2 -c_s^2(c_{i,x}c_{i,y}^2+c_{i,x}c_{i,z}^2)+c_s^4c_{i,x}, \nonumber \\
    M_{26 i} &=& c_{i,x}^2c_{i,y}^2c_{i,z}^2 -c_s^2(c_{i,x}^2c_{i,y}^2+c_{i,x}^2c_{i,z}^2+c_{i,y}^2c_{i,z}^2)+\nonumber \\
    & &c_s^4(c_{i,x}^2+c_{i,y}^2+c_{i,z}^2)-c_s^6. \nonumber 
\end{eqnarray}
The corresponding matrix $\mathbf{T}$ can be evaluated by replacing $c_{i, \alpha}$ by $\bar{c}_{i, \alpha}$. With this basis, relaxation frequencies are $\Lambda_{0-6}=\Lambda_{10-26}=1$ and $\Lambda_{7-9}=\omega$, while $\tilde{I}_{0-3}=0$ and $\tilde{I}_{4-26}=1$. Coefficients $b_v$ follow directly from the weighted orthogonality relation, that is
\[
\sum_i w_i T_{vi} T_{\ell i} = b_v \delta_{v\ell}
\]
and are reported below:
\begin{align}
&b_0  = 1, \quad b_{1-3} = \frac{1}{3}, \quad b_{4-6} = \frac{2}{9}, \nonumber \\
&b_{7-9}  = \frac{1}{9}, \quad b_{10-15} = \frac{2}{27}, \quad b_{16} = \frac{1}{27}, \nonumber \\
&b_{17-19} = \frac{4}{81}, \quad b_{20-22} = \frac{2}{81}, \quad b_{23-25} = \frac{4}{243}, \quad b_{26}  = \frac{8}{729}. \nonumber
\end{align}
Equilibrium central moments are all zero, except for $k_0^{eq}=\rho$. These properties are a direct consequence of the choice of a suitable basis which is applied to equilibrium distributions written up to the maximum order allowed by the present lattice discretisation~\cite{derosisHermite}. The post-collision state in terms of central moments is
\begin{eqnarray}
    k_0^{\star} &=& \rho, \\
    k_1^{\star} &=& 0, \nonumber \\
    k_2^{\star} &=& 0, \nonumber \\
    k_3^{\star} &=& 0, \nonumber \\
    k_4^{\star} &=& \sqrt{2 \rho \, k_B \, T \,c_s^2} \, \eta_4, \nonumber \\  
    k_5^{\star} &=& \sqrt{2 \rho \, k_B \, T\, c_s^2} \, \eta_5, \nonumber \\  
    k_6^{\star} &=& \sqrt{2 \rho \, k_B \, T \,c_s^2} \, \eta_6, \nonumber \\
    k_7^{\star} &=& (1-\omega)k_7 + \sqrt{\rho \, k_B \, T \,c_s^2 \,\omega (2-\omega)} \, \eta_7, \nonumber \\
    k_8^{\star} &=& (1-\omega)k_8 + \sqrt{\rho \, k_B \, T \,c_s^2 \,\omega (2-\omega)} \, \eta_8, \nonumber \\
    k_9^{\star} &=& (1-\omega)k_9 + \sqrt{\rho \, k_B \, T\, c_s^2 \,\omega (2-\omega)}  \,\eta_9, \nonumber \\
    k_{10}^{\star} &=& \sqrt{2 \rho \, k_B \, T } \,c_s^2  \,\eta_{10}, \nonumber \\
    k_{11}^{\star} &=& \sqrt{2 \rho \, k_B \, T }  \,c_s^2 \, \eta_{11}, \nonumber \\
    k_{12}^{\star} &=& \sqrt{2 \rho \, k_B \, T }  \,c_s^2  \,\eta_{12}, \nonumber \\
    k_{13}^{\star} &=& \sqrt{2 \rho \, k_B \, T } \, c_s^2 \, \eta_{13}, \nonumber \\
    k_{14}^{\star} &=& \sqrt{2 \rho \, k_B \, T }  \,c_s^2 \, \eta_{14}, \nonumber \\
    k_{15}^{\star} &=& \sqrt{2 \rho \, k_B \, T } \, c_s^2 \, \eta_{15}, \nonumber \\
    k_{16}^{\star} &=& \sqrt{\rho \, k_B \, T } \, c_s^2 \, \eta_{16}, \nonumber \\
    k_{17}^{\star} &=& 2\sqrt{\rho \, k_B \, T} \, c_s^3  \, \eta_{17}, \nonumber \\
    k_{18}^{\star} &=& 2\sqrt{\rho \, k_B \, T} \, c_s^3  \, \eta_{18}, \nonumber \\
    k_{19}^{\star} &=& 2\sqrt{\rho \, k_B \, T} \, c_s^3  \, \eta_{19}, \nonumber \\
    k_{20}^{\star} &=& \sqrt{2 \rho \, k_B \, T} \, c_s^3  \, \eta_{20}, \nonumber \\
    k_{21}^{\star} &=& \sqrt{2 \rho \, k_B \, T} \, c_s^3  \, \eta_{21}, \nonumber \\
    k_{22}^{\star} &=& \sqrt{2 \rho \, k_B \, T } \, c_s^3  \, \eta_{22}, \nonumber \\
    k_{23}^{\star} &=& 2\sqrt{\rho \, k_B \, T }  \, c_s^4 \, \eta_{23}, \nonumber \\
    k_{24}^{\star} &=& 2\sqrt{\rho \, k_B \, T } \, c_s^4 \, \eta_{24}, \nonumber \\
    k_{25}^{\star} &=& 2\sqrt{\rho \, k_B \, T } \, c_s^4  \, \eta_{25}, \nonumber \\
    k_{26}^{\star} &=& \sqrt{8\rho \, k_B \, T } \, c_s^5 \, \eta_{26}, \nonumber 
\end{eqnarray}
where prefactors depend on the values of $b_v$ and $\Lambda_v$ in Eq.~(\ref{eq:phi_general}). Relevant pre-collision central moments are computed as
\begin{eqnarray}
    k_7 &=& \sum_i f_i \, \bar{c}_{i,x} \bar{c}_{i,y} =  \nonumber \\
    & & f_7 - f_8 - f_9 + f_{10} + f_{19} - f_{20} - f_{21} + \nonumber \\
    & & f_{22} + f_{23} - f_{24} - f_{25} + f_{26} - \rho u_x u_y, \\
    k_8 &=&  \sum_i f_i \, \bar{c}_{i,x} \bar{c}_{i,z} =  \nonumber \\
    & & f_{11} - f_{12} - f_{13} + f_{14} + f_{19} - f_{20} + f_{21} -  \nonumber \\
    & &f_{22} - f_{23} + f_{24} - f_{25} + f_{26}  - \rho u_x u_z, \nonumber \\
    k_9 &=&   \sum_i f_i \, \bar{c}_{i,y} \bar{c}_{i,z} =  \nonumber \\
    & & f_{15}- f_{16} - f_{17} + f_{18} + f_{19} + f_{20} - f_{21} - \nonumber \\
    & & f_{22} - f_{23} - f_{24} + f_{25} + f_{26} - \rho u_y u_z, \nonumber
\end{eqnarray}
which correspond to the three independent off-diagonal components of the second-order central-moment tensor.\\
\indent Post-collision raw moments are available by Eq.~(\ref{eq:post_coll_rms_compact}), which allows us to reconstruct post-collision populations by Eq.~(\ref{eq:post_coll_pops_compact}). These are eventually streamed by Eq.~(\ref{streaming}). The resulting expressions are algebraically lengthy. For completeness and reproducibility, we therefore provide two MATLAB scripts (D2Q9-CM-FLBM.m and D3Q27-CM-FLBM.m) that carry out all symbolic manipulations required to construct the present models. It is worth to highlight that the present formulation naturally embeds within the general multiple-relaxation-time framework of central-moments-based LBMs. Setting $\mathbf{T}=\mathbf{M}$ reduces the scheme to its raw-moments counterpart, with $\mathbf{N}=\mathbf{I}$, while the BGK model is recovered by enforcing a uniform relaxation rate $\boldsymbol{\Lambda}=\omega\, \mathbf{I}$. The interested reader can also use the MATLAB scripts to test different basis (orthogonal and non-orthogonal), collision operators (central-moments- and raw-moments-based) and equilibrium truncation order (from the classical second-order one up to the maximum allowed by the given lattice discretisation).

\subsubsection*{Lattice-agnostic construction and the D3Q19 example}

Although the present formulation is illustrated primarily on the D3Q27 lattice, the underlying construction is not tied to a specific velocity set. The essential ingredients are (i) an orthogonal basis of central momebnts spanning the conserved and non-hydrodynamic subspaces and (ii) an equilibrium distribution constructed consistently with that basis. When these conditions are satisfied, the equilibrium manifold in CMs space retains the remarkably simple form
$\displaystyle \mathbf{k}^{eq} = \big(\rho,0,0,\ldots\big)^{\mathsf T}$, that is, only the zeroth-order central moment is populated at equilibrium, while all higher-order central moments vanish identically. This property is crucial for the fluctuating formulation: the absence of deterministic equilibrium contributions in non-conserved moments yields a clean separation between equilibrium structure and stochastic forcing, thereby simplifying both the construction of the noise covariance and its consistent projection onto the kinetic subspace. Thermodynamic consistency and fluctuation--dissipation balance can therefore be preserved on reduced lattices, provided that the equilibrium and the moment basis are designed in a compatible manner.

To make this point concrete, we report below an explicit  construction on the D3Q19 lattice, including the adopted basis and a fourth-order equilibrium distribution. The purpose is to demonstrate that the diagonal equilibrium structure discussed in \S\,\ref{sec:D3Q27} can be recovered on a reduced velocity set.

The D3Q19 lattice is defined by the discrete velocities
\begin{equation}
\bm{c}_i=(c_{i,x},c_{i,y},c_{i,z})\in\{-1,0,1\}^3,
\end{equation}
Explicitly, we have
\begin{align}
(c_{i,x},c_{i,y},c_{i,z}) = 
&(0,0,0), \nonumber\\
&(\pm1,0,0),(0,\pm1,0),(0,0,\pm1), \nonumber\\
&(\pm1,\pm1,0),(\pm1,0,\pm1),(0,\pm1,\pm1).
\end{align}
The corresponding lattice weights are
\begin{equation}
w_i=
\begin{cases}
1/3,  & \bm{c}_i=(0,0,0),\\[2pt]
1/18, & |\bm{c}_i|^2=1,\\[2pt]
1/36, & |\bm{c}_i|^2=2.
\end{cases}
\end{equation}
For notational convenience, let us define the velocity invariant
\begin{equation}
c^2 = c_{i,x}^2+c_{i,y}^2+c_{i,z}^2 .
\end{equation}

The central-moment basis employed in this work is specified by the polynomial set~\cite{PhysRevE.76.036704}
\begin{align}
M_{0,i}  &= 1, \\
M_{1,i}  &= c_{i,x}, \nonumber\\
M_{2,i}  &= c_{i,y},  \nonumber\\
M_{3,i}  &= c_{i,z}, \nonumber\\
M_{4,i}  &= c^2-1, \nonumber\\
M_{5,i}  &= 3c_{i,x}^2-c^2, \nonumber\\
M_{6,i}  &= c_{i,y}^2-c_{i,z}^2, \nonumber\\
M_{7,i}  &= c_{i,x}c_{i,y},  \nonumber\\
M_{8,i}  &= c_{i,y}c_{i,z}, \nonumber\\
M_{9,i}  &= c_{i,x}c_{i,z}, \nonumber\\
M_{10,i} &= c_{i,x}(3c^2-5), \nonumber\\
M_{11,i} &= c_{i,y}(3c^2-5), \nonumber\\
M_{12,i} &= c_{i,z}(3c^2-5), \nonumber\\
M_{13,i} &= c_{i,x}(c_{i,y}^2-c_{i,z}^2), \nonumber\\
M_{14,i} &= c_{i,y}(c_{i,z}^2-c_{i,x}^2), \nonumber\\
M_{15,i} &= c_{i,z}(c_{i,x}^2-c_{i,y}^2), \nonumber\\
M_{16,i} &= 3c^4-6c^2+1, \nonumber\\
M_{17,i} &= (2c^2-3)(3c_{i,x}^2-c^2), \nonumber\\
M_{18,i} &= (2c^2-3)(c_{i,y}^2-c_{i,z}^2).
\end{align}
The corresponding central-moment transformation matrix $\mathbf{T}$ is obtained by replacing $\bm{c}_i$ by $\bar{\bm{c}}_i=\bm{c}_i-\bm{u}$. The modal norms $b_v$ associated with the weighted orthogonality relation are
\begin{eqnarray}
b_0 &=& 1,\\
b_1 &=& 1/3, \nonumber\\
b_2 &=& 1/3, \nonumber\\
b_3 &=& 1/3, \nonumber\\
b_4 &=& 2/3, \nonumber\\
b_5 &=& 4/3, \nonumber\\
b_6 &=& 4/9, \nonumber\\
b_7 &=& 1/9, \nonumber\\
b_8 &=& 1/9, \nonumber\\
b_9 &=& 1/9, \nonumber\\
b_{10} &=& 2/3, \nonumber\\
b_{11} &=& 2/3, \nonumber\\
b_{12} &=& 2/3, \nonumber\\
b_{13} &=& 2/9, \nonumber\\
b_{14} &=& 2/9, \nonumber\\
b_{15} &=& 2/9, \nonumber\\
b_{16} &=& 2, \nonumber\\
b_{17} &=& 4/3, \nonumber\\
b_{18} &=& 4/9. \nonumber
\end{eqnarray}

We employ equilibrium populations that differ slightly from those reported by De Rosis \& Coreixas~\cite{de2020multiphysics}. In the present case, the rest population reads
\begin{equation}
f_0^{eq}
=
\rho\left(
w_0+\frac{1}{6}u^2+\frac{1}{2}u^4,
\right).
\end{equation}
with $u^2=u_x^2+u_y^2+u_z^2$. For the axis populations, characterised by a single non-zero component $c_{i,\alpha}=\pm 1$ with $\alpha\in\{x,y,z\}$, we use the unified expression
\begin{equation}
f_i^{eq}
=
\rho\left(
w_i+\frac{1}{6}c_{i,\alpha}u_\alpha
-\frac{1}{3}u^2
-\frac{1}{2}c_{i,\alpha}u_\alpha u^2
+\frac{1}{2}u^4
\right),
\end{equation}
with the remaining velocity components vanishing.
Finally, for the face-diagonal populations with exactly two non-zero components $c_{i,\alpha}=\pm 1$, $c_{i,\beta}=\pm 1$ and $c_{i,\gamma}=0$, we write
\begin{equation}
\begin{aligned}
f_i^{eq}
=
\rho\Bigg[
&w_i
+\frac{1}{12}\left(c_{i,\alpha}u_\alpha+c_{i,\beta}u_\beta\right)
+\frac{5}{24}\left(u_\alpha^2+u_\beta^2\right) - \\
&\frac{1}{8}u_\gamma^2 +\frac{1}{4}c_{i,\alpha}c_{i,\beta}u_\alpha u_\beta 
+\frac{1}{8}\left(c_{i,\alpha}u_\alpha^3+c_{i,\beta}u_\beta^3\right)+\\
&\frac{1}{4}\left(c_{i,\alpha}u_\alpha u_\beta^2+c_{i,\beta}u_\beta u_\alpha^2\right)- \\
&\frac{1}{8}\left(u_\alpha^4+u_\beta^4\right) 
+\frac{1}{4}u_\alpha^2u_\beta^2
+\frac{3}{8}u_\gamma^4
\Bigg],
\end{aligned}
\end{equation}
where $(\alpha,\beta,\gamma)$ denotes any permutation of $(x,y,z)$ consistent with the lattice direction. Together, these expressions provide a complete fourth-order equilibrium for all 19 discrete velocities.

Projecting this equilibrium onto the above central-moment basis yields the equilibrium vector
\begin{equation}
\mathbf{k}^{eq} = \big(\rho,0,0,\ldots,0\big)^{\mathsf T},
\end{equation}
so that only the density mode is populated at equilibrium and all higher-order central moments vanish identically in the local rest frame. As discussed earlier in the manuscript, this structure enables a separation between deterministic equilibrium contributions and stochastic forcing and facilitates the consistent enforcement of the fluctuation--dissipation balance.

The post-collision state in central-moment space follows by selecting $\Lambda_{7-9}=\omega$ (with all remaining relaxation parameters equal to $1$), $\tilde{I}_{0-3}=0$ and $\tilde{I}_{4-18}=1$, yielding
\begin{eqnarray}
k_0^{\star} &=& \rho, \\
k_1^{\star} &=& k_2^{\star} = k_3^{\star} = 0 , \nonumber \\
k_4^{\star} &=& \sqrt{2 \rho \, k_B \, T}\, \eta_4 , \nonumber \\
k_5^{\star} &=& 2\sqrt{\rho \, k_B \, T} \,\eta_5, \nonumber \\
k_6^{\star} &=& 2\sqrt{\rho \, k_B \, T \, c_s^2} \,\eta_6, \nonumber \\
k_7^{\star} &=& (1-\omega)k_7 + \sqrt{\rho \, k_B \, T\,c_s^2 \, \omega (2-\omega)} \,\eta_7, \nonumber \\
k_8^{\star} &=& (1-\omega)k_8 + \sqrt{\rho \, k_B \, T\,c_s^2 \, \omega (2-\omega)} \,\eta_8, \nonumber \\
k_9^{\star} &=& (1-\omega)k_9 + \sqrt{\rho \, k_B \, T\,c_s^2 \, \omega (2-\omega)}\,\eta_9, \nonumber \\
k_{10}^{\star} &=& \sqrt{2 \rho \, k_B \, T} \,\eta_{10} , \nonumber \\
k_{11}^{\star} &=& \sqrt{2 \rho \, k_B \, T} \,\eta_{11} , \nonumber \\
k_{12}^{\star} &=& \sqrt{2 \rho \, k_B \, T} \,\eta_{12} , \nonumber \\
k_{13}^{\star} &=& \sqrt{2 \rho \, k_B \, T\,c_s^2} \,\eta_{13} , \nonumber \\
k_{14}^{\star} &=& \sqrt{2 \rho \, k_B \, T\,c_s^2}\, \eta_{14} , \nonumber \\
k_{15}^{\star} &=& \sqrt{2 \rho \, k_B \, T\,c_s^2}\, \eta_{15} , \nonumber \\
k_{16}^{\star} &=& \sqrt{2 \rho \, k_B \, T\,c_s^{-2}} \,\eta_{16} , \nonumber \\
k_{17}^{\star} &=& 2\sqrt{\rho \, k_B \, T}\, \eta_{17} , \nonumber \\
k_{18}^{\star} &=& 2\sqrt{\rho \, k_B \, T \, c_s^2} \,\eta_{18}. \nonumber
\end{eqnarray}
Post-collision raw moments are obtained as $\mathbf{r}^{\star}=\mathbf{N}^{-1}\mathbf{k}^{\star}$, and the post-collision populations follow as $\mathbf{f}^{\star}=\mathbf{M}^{-1}\mathbf{r}^{\star}$ prior to streaming. The full expressions are provided by the MATLAB script \texttt{D3Q19-CM-FLBM.m} in the Supplemental Material.

This explicit construction demonstrates that the equilibrium structure is not a peculiarity of D3Q27, but a structural consequence of pairing an orthogonal central-moment basis with a compatible equilibrium distribution. The fluctuating central-moment framework is therefore lattice-agnostic; analogous behaviour is expected on other reduced lattices such as D3Q15.

\section{Results and discussion}
The present fluctuating central-moments-based lattice Boltzmann methods (CM-FLBMs) are assessed through a sequence of seven systematic tests that probe deterministic consistency, thermodynamic correctness, and numerical robustness across a wide range of parameters, as summarised in Table~\ref{tab:validation_cases}.
\begin{table*}[t]
\centering
\caption{Summary of validation cases used to assess the fluctuating lattice Boltzmann solver.}
\label{tab:validation_cases}
\begin{tabular}{c|l|l|l|l}
\hline
\hline
Test & Purpose & Varied parameter(s) & Fixed parameter(s) & Expected behaviour \\
\hline
1 &
Zero-noise regression &
\(k_B T = 0\) &
\(\rho_0, \tau\) &
\(\langle u_x^2 \rangle = \langle u_y^2 \rangle = 0\) \\

2 &
Taylor--Green vortex &
\(k_B T = 0\) &
\(\rho_0, \tau\) &
Velocity decay according to analytical predictions \\

3 &
Equipartition &
-- &
\(k_B T, \rho_0, \tau\) &
\(\langle u_\alpha^2 \rangle = k_B T/\rho_0\) \\

4 &
Thermal scaling &
\(k_B T\) &
\(\rho_0, \tau\) &
\(\langle u_\alpha^2 \rangle \propto k_B T\) \\

5 &
Density scaling &
\(\rho_0\) &
\(k_B T, \tau\) &
\(\langle u_\alpha^2 \rangle \propto 1/\rho_0\) \\

6 &
Relaxation-time sweep &
\(\tau\) &
\(k_B T, \rho_0\) &
Equipartition independent of \(\tau\) \\

7 &
Anisotropic domain &
Aspect ratio &
\(k_B T, \rho_0, \tau\) &
Isotropic equipartition independent of geometry \\
\hline
\hline
\end{tabular}
\end{table*}

All tests are performed on periodic domains and compared against theoretical expectations derived from equilibrium statistical mechanics. Thermal fluctuations are introduced through stochastic forcing of the non-conserved central-moment kinetic modes during the collision step, with noise amplitudes proportional to \(\sqrt{k_B T}\), in accordance with Landau--Lifshitz fluctuating hydrodynamics. A reference time scale is defined as
\begin{equation}
t_0 = \frac{N_y}{u_0},
\end{equation}
where \(u_0\) is a prescribed characteristic velocity, chosen to ensure a low-Mach-number regime and to define a macroscopic advective time scale consistent with the weakly compressible lattice Boltzmann framework, and $N_y$ is the number of lattice point in the (reference) vertical direction. Each simulation is advanced for several multiples of \(t_0\) in order to reach a statistically stationary state. Spatial averages of the squared velocity components, \(\langle u_x^2 \rangle\), \(\langle u_y^2 \rangle\), and \(\langle u_z^2 \rangle\), are monitored as primary diagnostics. At thermal equilibrium, the theoretical prediction from equipartition reads
\begin{equation}
\langle u_\alpha^2 \rangle = \frac{k_B T}{\rho_0},
\end{equation}
for each Cartesian component \(\alpha\), where $k_B \, T$ denotes the thermal energy per lattice site (unit lattice volume), consistently with the ideal-gas equilibrium population variance.

Unless otherwise stated, the fluid is initialised at rest with uniform density \(\rho_0=1\), all simulations employ the same relaxation parameter, that is \(\tau = 1/\omega = 0.8\), and the characteristic velocity is \(u_0 = 0.01\). For the first six tests, simulations carried out using the D2Q9 and D3Q27 models lead to overlapping results; therefore, only the two-dimensional results are reported in the following, with the domain size being $N_x \times N_y = 200 \times 200$ lattice sites.

\subsection{Test 1: Zero-noise (deterministic) regression}
As a baseline verification, simulations are performed with thermal fluctuations disabled ($k_B T = 0$). In this limit, the fluctuating lattice Boltzmann method must reduce exactly to its deterministic counterpart. In our runs, the macroscopic velocity field remains identically zero up to machine precision, with vanishing second-order moments $\langle u_x^2 \rangle$ and $\langle u_y^2 \rangle$. This confirms that stochastic forcing is correctly suppressed and that no spurious numerical noise is introduced by the implementation. The test also verifies that mass and momentum conservation are preserved in the absence of fluctuations. The interested reader can reproduce these results by running the C++ programs provided in the Supplemental Material.

\subsection{Test 2: Deterministic Taylor--Green vortex}
\begin{figure}[htbp]
    \centering
    \includegraphics[width=0.99\linewidth]{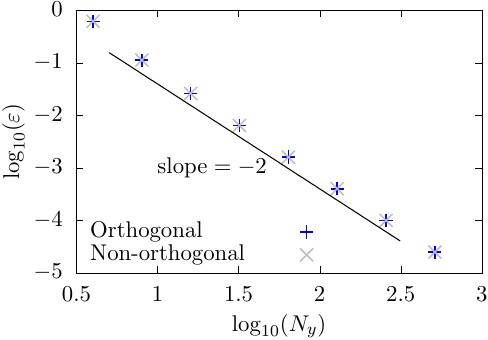}
    \caption{Test 2: grid-convergence of the Taylor--Green vortex. Log--log plot of the $L_2$-norm of the velocity error versus grid resolution, showing a fitted slope of $-2$, consistent with second-order accuracy of the scheme in space and time. Findings obtained by the adoption of an orthogonal basis (blue orthogonal crosses) overlap those generated by a non-orthogonal one (grey diagonal crosses).}
    \label{Figure1}
\end{figure}

As an additional deterministic validation, we consider the decay of a two-dimensional Taylor--Green vortex with thermal fluctuations disabled ($k_B T = 0$). The initial velocity field is prescribed as
\begin{eqnarray}
u_x(\bm{x}, t = 0) &=& -u_0 \cos(\kappa x)\sin(\kappa y), \\
u_y(\bm{x}, t = 0) &=& \;\;u_0 \sin(\kappa x)\cos(\kappa y), \nonumber
\end{eqnarray}
where $\kappa = 2\pi/N_y$. The reference velocity is set to $u_0 = 0.001$, and the Reynolds number is $\mathrm{Re} = u_0 N_y/\nu = 1000$. For an incompressible viscous flow, this configuration admits an analytical solution in which the velocity amplitude decays exponentially as $\exp(-2\nu \kappa^2 t)$.\\
\indent Numerical simulations are performed using the present formulation, with the kinematic viscosity set consistently with the relaxation frequency. The numerical velocity field is monitored and compared against the analytical prediction. Results from a grid-convergence analysis are reported in Figure~\ref{Figure1}, where the $L_2$-norm of the relative error between the numerical solution $\bm{u}$ and the analytical one $\bm{u}_{\mathrm{an}}$ is computed as
\begin{equation}
\varepsilon = \frac{\| \bm{u} - \bm{u}_{\mathrm{an}} \|}{\| \bm{u}_{\mathrm{an}} \|}.
\end{equation}
Grid refinement studies confirm that the error in the decay of the flow field exhibits second-order convergence with respect to spatial resolution, as expected for the underlying lattice Boltzmann discretisation. This test further demonstrates that, in the absence of thermal noise, the fluctuating formulation reduces exactly to its deterministic counterpart and reproduces the correct viscous dynamics with the expected order of accuracy.
\\
\indent For completeness, we repeat this test using a non-orthogonal basis of central moments~\cite{derosis2016epl_d2q9}. In the deterministic setting considered here, the resulting numerical solution coincides exactly with that obtained using the orthogonal basis. This observation warrants closer inspection, as the two formulations are not equivalent at the level of central moments.\\
\indent Indeed, when a non-orthogonal basis is employed, several post-collision central moments assume non-zero equilibrium values, namely
\begin{eqnarray}
k_0^{\star} &=& \rho, \\
k_3^{\star} &=& 2 \rho c_s^2, \nonumber \\
k_4^{\star} &=& (1-\omega) k_4, \nonumber \\
k_5^{\star} &=& (1-\omega) k_5, \nonumber \\
k_8^{\star} &=& \rho c_s^4, \nonumber
\end{eqnarray}
whereas in the orthogonal formulation one has $k_3^{\star}=k_8^{\star}=0$ (in the absence of stochastic forcing). The corresponding post-collision raw moments therefore differ between the two formulations and are given, in the non-orthogonal case, by
\begin{eqnarray}
    r_0^{\star} &=& k_0^{\star}, \\
    r_1^{\star} &=& \rho u_x, \nonumber \\
    r_2^{\star} &=& \rho u_y, \nonumber \\
    r_3^{\star} &=& \rho (u_x^2 + u_y^2 + 2 c_s^2), \nonumber \\
    r_4^{\star} &=& k_4^{\star} + \rho (u_x^2 - u_y^2), \nonumber \\
    r_5^{\star} &=& k_5^{\star} + \rho u_x u_y, \nonumber \\
    r_6^{\star} &=& 2 u_x k_5^{\star} + \frac{1}{2}u_y (2 \rho c_s^2 + k_4^{\star}) + \rho u_x^2 u_y , \nonumber \\
    r_7^{\star} &=& 2 k_5^{\star} u_y + \frac{1}{2} u_x (2 \rho c_s^2- k_4^{\star}) + \rho u_x u_y^2 , \nonumber \\
    r_8^{\star} &=& 4 u_x u_y k_5^{\star}  - \frac{1}{2}  (u_x^2 - u_y^2) k_4^{\star} + \rho [u_x^2 u_y^2+c_s^4+c_s^2 (u_x^2 + u_y^2)], \nonumber 
\end{eqnarray}
while the orthogonal formulation yields
\begin{eqnarray}
    r_0^{\star} &=& k_0^{\star}, \\
    r_1^{\star} &=& \rho  u_x, \nonumber \\
    r_2^{\star} &=& \rho  u_y, \nonumber \\
    r_3^{\star} &=& \rho(u_x^2+u_y^2), \nonumber \\
    r_4^{\star} &=& k_4^{\star}+\rho (u_x^2-u_y^2), \nonumber \\
    r_5^{\star} &=& k_5^{\star}+\rho u_x u_y, \nonumber \\
    r_6^{\star} &=& 2u_y k_5^{\star} + \frac{1}{2}u_y k_4^{\star} + \rho u_x^2 u_y, \nonumber \\
    r_7^{\star} &=& 2u_y k_5^{\star} - \frac{1}{2} u_x k_4^{\star} + \rho u_x u_y^2, \nonumber \\
    r_8^{\star} &=&  4 u_x u_y k_5^{\star} -\frac{1}{2}(u_x^2-u_y^2) k_4^{\star} + \rho u_x^2 u_y^2.\nonumber 
\end{eqnarray}
\\
\indent Post-collision population in the non-orthogonal case read as follows:
\begin{eqnarray}
    f_0^{\star} &=& r_0^{\star} - r_3^{\star} + r_8^{\star}, \\
    f_1^{\star} &=& \frac{r_3^{\star} + r_4^{\star}}{4} + \frac{r_1^{\star} - r_7^{\star} - r_8^{\star}}{2}, \nonumber \\
    f_2^{\star} &=& \frac{r_3^{\star} - r_4^{\star}}{4} + \frac{r_2^{\star} - r_6^{\star} - r_8^{\star}}{2}, \nonumber \\
    f_3^{\star} &=& \frac{r_3^{\star} + r_4^{\star}}{4} + \frac{-r_1^{\star} + r_7^{\star} - r_8^{\star}}{2}, \nonumber \\
    f_4^{\star} &=& \frac{r_3^{\star} - r_4^{\star}}{4} + \frac{-r_2^{\star} + r_6^{\star} - r_8^{\star}}{2}, \nonumber \\
    f_5^{\star} &=& \frac{r_5^{\star} + r_6^{\star} + r_7^{\star} + r_8^{\star}}{4}, \nonumber \\
    f_6^{\star} &=& \frac{-r_5^{\star} + r_6^{\star} - r_7^{\star} + r_8^{\star}}{4}, \nonumber \\
    f_7^{\star} &=& \frac{r_5^{\star} - r_6^{\star} - r_7^{\star} + r_8^{\star}}{4}, \nonumber \\
    f_8^{\star} &=& \frac{-r_5^{\star} - r_6^{\star} + r_7^{\star} + r_8^{\star}}{4}.\nonumber
\end{eqnarray}
\\
\indent At first sight, these expressions appear substantially different. However, upon reconstructing the post-collision populations from the corresponding raw and central moments, all additional equilibrium contributions introduced by the non-orthogonal basis cancel exactly. As a result, the post-collision populations coincide with those obtained using the orthogonal formulation. For instance, the rest population can be written as
\begin{equation}
\begin{aligned}
f_0^{\star} ={}& \rho \left[ u_x^2 u_y^2-2 c_s^2\left(u_x^2+u_y^2\right) + 4c_s^4 \right] + \\
& \frac{u_y^2-u_x^2}{2} (1-\omega) k_4
+ 4 u_x u_y (1-\omega) k_5,
\end{aligned}
\end{equation}
independently of whether an orthogonal or non-orthogonal central-moment basis is adopted.\\
\indent This equivalence is specific to the deterministic setting considered here and arises for other lattice discretisations. While non-orthogonal bases yield identical macroscopic behaviour in the absence of fluctuations, they fundamentally differ from orthogonal formulations once stochastic forcing is introduced. In particular, the presence of non-zero equilibrium values in higher-order central moments necessitates correlated noise to enforce fluctuation--dissipation balance, whereas orthogonal formulations admit diagonal equilibrium covariance and independent stochastic forcing by construction. This distinction underpins the fluctuating formulation developed in the present work.

\subsection{Test 3: Equipartition at equilibrium}
\begin{figure}[htbp]
    \centering
    \includegraphics[width=0.99\linewidth]{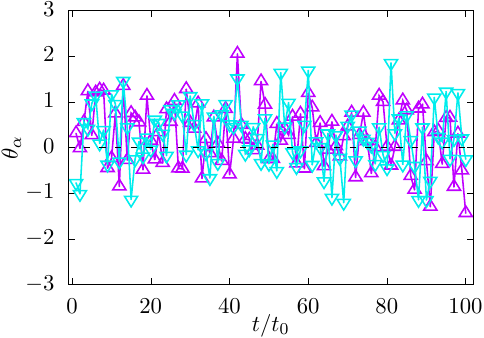}
    \caption{Test 3. Time evolution of the domain-averaged velocity variances $\langle u_x^2\rangle$ (dark magenta line with triangles) and $\langle u_y^2\rangle$ (dark cyan line with inverted triangles) for a homogeneous system with thermal fluctuations enabled ($k_B T = 1/3000$, $\rho_0=1$). The black dashed line indicates the equipartition prediction. Both velocity components fluctuate around the expected value, confirming correct calibration of the stochastic forcing.}
    \label{Figure2}
\end{figure}

This case verifies that the fluctuating solver relaxes toward a thermodynamic equilibrium state consistent with equipartition of kinetic energy. Thermal fluctuations are activated with a fixed thermal energy
\begin{equation}
k_B T = \frac{1}{3000},
\end{equation}
while the density and relaxation parameters are held constant. The system is evolved from rest until a statistically stationary state is reached. At equilibrium, the mean velocities are expected to vanish,
\begin{equation}
\langle u_x \rangle \approx \langle u_y \rangle \approx 0,
\end{equation}
while the velocity variances converge to the equipartition value,
\begin{equation}
\langle u_x^2 \rangle \approx \langle u_y^2 \rangle \approx \frac{k_B T}{\rho_0},
\end{equation}
with isotropy between the two velocity components.\\
\indent Statistics are collected over $T_{\mathrm{max}} = 100 t_0$. To quantify deviations from equipartition in a dimensionless and component-wise manner, we introduce the relative error
\begin{equation}
\theta_{\alpha}= \frac{\dfrac{k_B T}{\rho_0}-\langle u_{\alpha}^2 \rangle}
     {\dfrac{k_B T}{\rho_0}} \times 100,
\end{equation}
which measures the percentage deviation of the measured second-order velocity moment from the theoretical target imposed by fluctuating hydrodynamics. Figure~\ref{Figure2} reports the temporal evolution of $\theta_{\alpha}$ for both velocity components. The error remains tightly distributed around zero over the entire observation window, with fluctuations bounded within approximately $\pm 2\%$. No systematic drift or long-time trend is observed, indicating that the fluctuating scheme preserves equipartition on average. The symmetric distribution of $\theta_{\alpha}$ about zero further confirms the absence of directional bias between velocity components, while the residual oscillations reflect finite-sample statistical fluctuations rather than a structural deficiency of the model.
\\
\indent Moreover, deviations from equipartition are quantified using both signed mean errors (bias) and root-mean-square (RMS) errors. For each velocity component, the signed mean deviation is defined as
\begin{eqnarray}
\overline{\Delta u_x^2}
&=& \frac{1}{T_{\mathrm{max}}}\sum_t^{T_{\mathrm{max}}}
\left(\langle u_x^2\rangle(t) - \frac{k_B T}{\rho_0} \right), \\
\overline{\Delta u_y^2}
&=& \frac{1}{T_{\mathrm{max}}}\sum_t^{T_{\mathrm{max}}}
\left(\langle u_y^2\rangle(t) - \frac{k_B T}{\rho_0}\right),
\end{eqnarray}
while the isotropic kinetic-energy measure is given by
\begin{equation}
\overline{\Delta e}
= \frac{1}{T_{\mathrm{max}}}\sum_t^{T_{\mathrm{max}}}
\left(
\frac{\langle u_x^2\rangle(t) + \langle u_y^2\rangle(t)}{2}
- \frac{k_B T}{\rho_0}
\right).
\end{equation}
Corresponding RMS errors are defined as
\begin{equation}
\mathrm{RMS}(u_x^2)
=
\left[
\frac{1}{T_{\mathrm{max}}}\sum_t^{T_{\mathrm{max}}}
\left(\langle u_x^2\rangle(t) - \frac{k_B T}{\rho_0}\right)^2
\right]^{1/2},
\end{equation}
with analogous expressions for $\mathrm{RMS}(u_y^2)$ and $\mathrm{RMS}(e)$. Quantitative measures of equipartition accuracy are summarised in Table~\ref{tab:equipartition_errors}. 
The measured mean signed deviations of the second-order velocity moments remain of order $10^{-7}$, corresponding to relative biases below $0.3\%$ of the theoretical target $k_B T/\rho_0$. 
Root-mean-square errors are of order $10^{-6}$, with the isotropic kinetic-energy measure exhibiting the smallest value, thus further confirming the absence of systematic anisotropy and the statistical consistency of the fluctuating formulation.
\begin{table}
\centering
\caption{Test 3: mean signed deviations (bias) and root-mean-square (RMS) errors of the second-order velocity moments with respect to the theoretical equipartition target $k_B T/\rho_0$.}
\label{tab:equipartition_errors}
\begin{tabular}{l|c|c}
\hline
\hline
Quantity & Mean bias & RMS error \\
\hline
$\langle u_x^2\rangle$ & $8.76\times10^{-7}$ & $2.4\times10^{-6}$ \\
$\langle u_y^2\rangle$ & $6.31\times10^{-7}$ & $2.5\times10^{-6}$ \\
$\langle (u_x^2+u_y^2)/2\rangle$ & $7.54\times10^{-7}$ & $1.8\times10^{-6}$ \\
\hline
\hline
\end{tabular}
\end{table}
\\
\indent These results demonstrate that the present fluctuating scheme reproduces kinetic-energy equipartition with high accuracy. Statistical fluctuations remain small and isotropic, and no systematic directional bias between velocity components is observed, confirming the consistency of the formulation with fluctuating hydrodynamics.

\subsection{Test 4: Scaling with thermal energy}
\begin{figure}[htbp]
    \centering
    \includegraphics[width=0.99\linewidth]{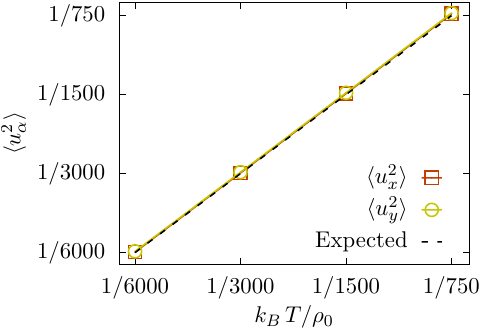}
    \caption{Test 4: scaling of the equilibrium velocity variance with thermal energy. Time-averaged values of $\langle u_x^2\rangle$ (dark orange line with squares) and $\langle u_y^2\rangle$ (dark yellow line with circles) as a function of $k_B T$ for a homogeneous system at equilibrium. The black dashed line indicates the equipartition prediction $\langle u_\alpha^2\rangle = k_B T/\rho_0$ ($\rho_0=1$). Both velocity components follow the expected linear scaling.}
    \label{Figure3}
\end{figure}

This case tests the linear dependence of velocity fluctuations on the thermal energy. A sequence of simulations is performed with
\begin{equation}
k_B T \in \left\{ \frac{1}{6000}, \frac{1}{3000}, \frac{1}{1500}, \frac{1}{750} \right\},
\end{equation}
while all other parameters, including \(\rho_0\) and \(\tau\), are kept fixed. Each simulation includes an initial warm-up phase to eliminate transient effects, followed by a long averaging interval. The expected outcome is a linear scaling of both \(\langle u_x^2 \rangle\) and \(\langle u_y^2 \rangle\) with \(k_B T\), with slope \(1/\rho_0\), confirming the thermodynamic consistency of the fluctuating collision operator.
\\
\indent The result in Figure~\ref{Figure3} confirms that the stochastic forcing enters the dynamics in a physically consistent manner and that no nonlinear artefacts arise from the discretisation or moment reconstruction.

\subsection{Test 5: Density scaling}
\begin{figure}[htbp]
    \centering
    \includegraphics[width=0.99\linewidth]{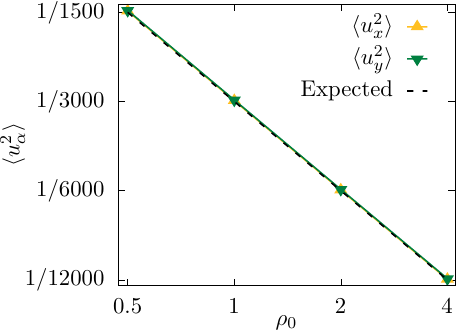}
    \caption{Test 5: scaling of the equilibrium velocity variance with density. Time-averaged values of $\langle u_x^2\rangle$ (goldenrod line with filled triangles) and $\langle u_y^2\rangle$ (dark spring green line with inverted filled triangles) as a function of the reference density $\rho_0$ for a homogeneous system at equilibrium, with fixed thermal energy $k_B T$. The black dashed line indicates the equipartition prediction $\langle u_\alpha^2\rangle = k_B T/\rho_0$. Both velocity components collapse onto the theoretical $k_B T/\rho_0$ scaling, demonstrating consistency with the equipartition theorem.}
    \label{Figure4}
\end{figure}

In the fifth test, we assess the inverse dependence of velocity fluctuations on the mass density. The thermal energy is fixed at \(k_B T = 1/3000\), while the reference density is varied according to
\begin{equation}
\rho_0 \in \{0.5,\,1,\,2,\,4\}.
\end{equation}
At equilibrium, the velocity variance is expected to scale as
\begin{equation}
\langle u_\alpha^2 \rangle \sim \frac{k_B T}{\rho_0}.
\end{equation}
\indent The numerical results shown in Figure~\ref{Figure4} follow this scaling closely, demonstrating that the implementation correctly accounts for density dependence in the fluctuation amplitudes. This is a non-trivial verification, as an incorrect placement of density factors in the noise terms would immediately manifest in this test.

\subsection{Test 6: Relaxation-time sweep}
\begin{figure*}[!htbp]
    \centering
    \subfigure[]{\includegraphics[width=0.49\linewidth]{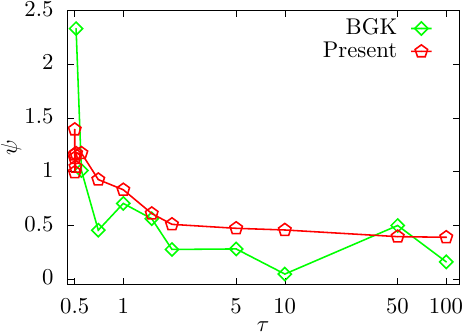} \label{Figure5a}}
    \subfigure[]{\includegraphics[width=0.49\linewidth]{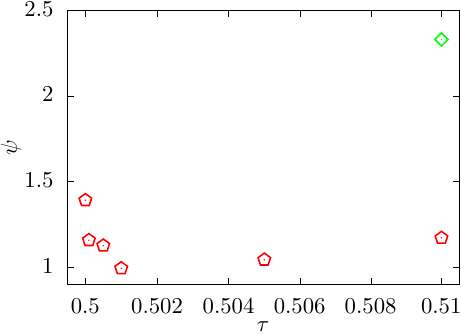} \label{Figure5b}}
    \caption{Test 6: relative deviation $\psi$ from the equipartition target as a function of the relaxation time $\tau$. (a) Full $\tau$--range comparison between the BGK scheme (green line with diamonds) and the present formulation (red line with pentagons). While both methods recover comparable accuracy at moderate and large $\tau$, BGK exhibits a sharp increase in error as $\tau \to 0.5$. (b) Zoomed view near the stability limit $\tau=0.5$, showing the abrupt numerical breakdown of BGK, characterized by large excursions and the onset of NaN values, in contrast to the smooth and robust behaviour of the present method.}
    \label{Figure5}
\end{figure*}

This case examines the robustness of the fluctuating formulation with respect to changes in the relaxation time and verifies that equilibrium velocity fluctuations are independent of viscosity. The thermal energy and reference density are kept fixed at
\[
k_B T = \frac{1}{3000}, \qquad \rho_0 = 1,
\]
while the relaxation time \(\tau\) is varied over a wide rangefrom near the stability limit $\tau \to 0.5$ up to highly diffusive regimes $\tau \gg 1$. Specifically, the following values of the relaxation time are considered~\cite{10.1063/5.0288232}:
\begin{eqnarray}
\tau &\in &\{0.5,\; 0.5001,\; 0.5005,\; 0.501,\; 0.505,\; 0.51,\\
           & &0.55,\; 0.7,\; 1.0,\; 1.5,\; 2.0,\; 5.0,\; 10.0,\; 50.0,\; 100.0\}. \nonumber
\end{eqnarray}
The corresponding kinematic viscosity is adjusted consistently through the relation \(\tau = \nu/c_s^2 + 1/2\). For each value of \(\tau\), the system is evolved from rest to a statistically stationary state. The spatially averaged velocity variances are then measured. Provided numerical stability is maintained, the expected outcome is that the equilibrium velocity fluctuations satisfy
\begin{equation}
\langle u_x^2 \rangle \approx \langle u_y^2 \rangle \approx \frac{k_B T}{\rho_0},
\end{equation}
independently of the relaxation time, thereby demonstrating that the thermodynamic equilibrium properties of the fluctuating scheme are viscosity-independent.\\
\indent In Figure~\ref{Figure5}, we plot 
\begin{equation}
    \psi = \frac{| \frac{k_B \, T}{\rho_0} - \frac{1}{2}\left( \langle u_x^2\rangle + \langle u_y^2\rangle \right)|}{\frac{k_B \, T}{\rho_0}} \times 100,
\end{equation}
that is the absolute relative deviation from the equipartition target as a function of the relaxation time $\tau$ for the standard BGK scheme and the present formulation. Over the entire range of $\tau$ shown in Figure~\ref{Figure5a}, the present method exhibits a weak and smooth dependence on $\tau$, maintaining uniformly small errors even in the strongly over--relaxed regime. By contrast, the BGK scheme displays a qualitatively different behaviour: as $\tau$ is reduced towards the stability limit $\tau = 0.5$, the error increases sharply and the method undergoes a numerical breakdown. This breakdown is not a gradual deterioration of accuracy but a genuine loss of numerical well-posedness, manifested by the appearance of unbounded fluctuations and the occurrence of NaN values in the computed statistics.
\\
\indent The zoomed view in Figure~\ref{Figure5b} shows that this failure is abrupt and highly localised in parameter space, occurring in the immediate vicinity of $\tau = 0.5$. In this regime, BGK ceases to produce meaningful results, whereas the present formulation remains fully stable and continues to enforce equipartition without any indication of pathological behaviour. These results demonstrate that the present approach possesses a substantially enlarged domain of robustness compared with BGK and can be safely operated up to the over--relaxation limit where the classical scheme becomes ill-posed.

\subsection{Test 7: Anisotropic three-dimensional domain}
\begin{figure}[htbp]
    \centering
    \includegraphics[width=0.99\linewidth]{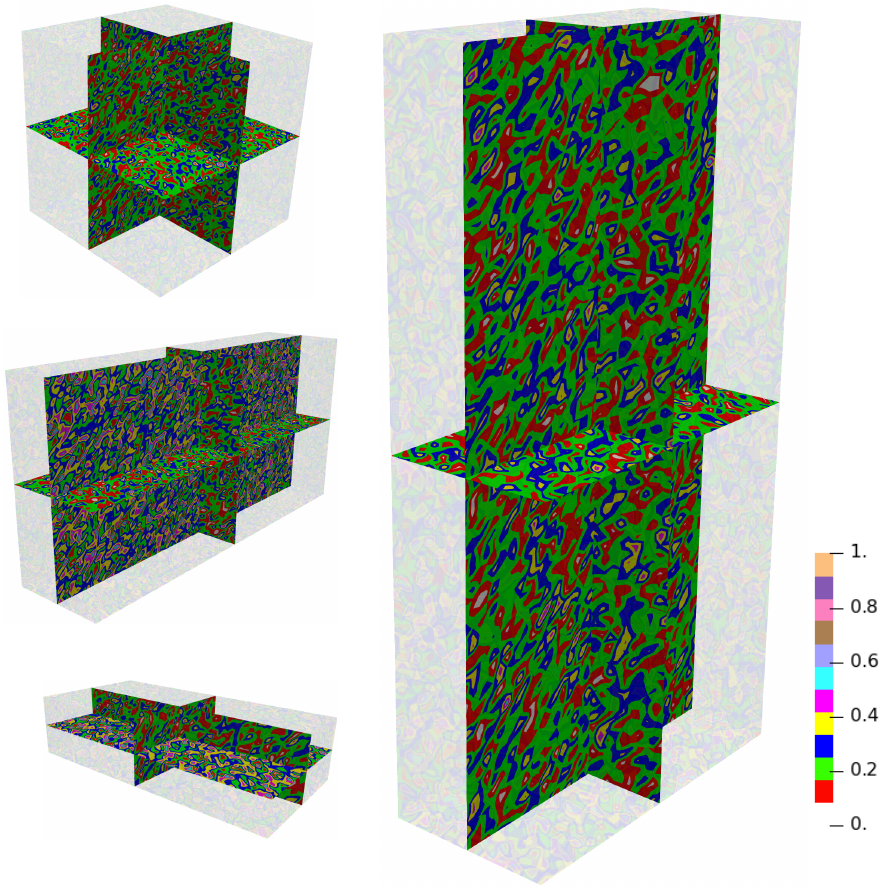}
    \caption{Test 7: instantaneous three-dimensional visualisation of the velocity field at equilibrium in an anisotropic periodic domain for all the scenarios (C1 - top left panel, C2 - mid left panel, C3 - bottom left panel, C4 - right panel). Coloured slices show the velocity field on mutually orthogonal planes intersecting the domain centre, illustrating the spatial structure induced by thermal fluctuations. The absence of coherent patterns or directional preference confirms statistical homogeneity and isotropy, with fluctuations uniformly distributed throughout the domain despite geometric anisotropy.}
    \label{Figure7}
\end{figure}

To assess the robustness of the fluctuating formulation in fully three-dimensional settings and to verify that geometric anisotropy does not compromise isotropy in velocity space, we perform equipartition tests at equilibrium on periodic domains with varying aspect ratios. The following configurations are considered:
\begin{itemize}
    \item C1: $(N_x,N_y,N_z)=(50,50,50)$;
    \item C2: $(N_x,N_y,N_z)=(25,50,100)$;
    \item C3: $(N_x,N_y,N_z)=(60,15,30)$;
    \item C4: $(N_x,N_y,N_z)=(20,80,40)$.
\end{itemize}
In a correct isotropic fluctuating scheme, the second-order velocity moments must remain isotropic even when the computational domain is strongly anisotropic, namely
\[
\langle u_x^2\rangle \approx \langle u_y^2\rangle \approx \langle u_z^2\rangle,
\]
with any residual differences decreasing as the sampling time increases. This test is particularly stringent, as anisotropic domains readily expose subtle implementation errors, such as direction-dependent streaming mistakes, index-ordering bugs, or unintended coupling between random-number generation and memory layout that may remain hidden in cubic configurations.\\
\indent Table~\ref{tab:equipartition_3D} summarises the equipartition statistics for all four cases in a compact form. For each configuration, we report the mean signed deviation (bias) and
root-mean-square (RMS) error of the total kinetic energy,
\[
e=\frac{1}{3}\sum_{\alpha}\langle u_\alpha^2\rangle,
\]
together with the maximum component-wise deviation from the mean bias, which provides a direct measure of residual anisotropy. Across all aspect ratios, the mean bias $\overline{\Delta e}$ remains of order $10^{-6}$, corresponding to relative deviations below $0.5\%$ of the equipartition target $k_B T/\rho_0$. The RMS errors are consistently below $1\%$, confirming that deviations from equipartition are weak and statistically bounded. The maximum component-wise deviation is more than an order of magnitude smaller than the mean bias in all cases, indicating that residual anisotropy
is negligible compared with the overall equipartition error.
\begin{table}[!htbp]
\centering
\caption{Test 7: summary of equipartition statistics for anisotropic three-dimensional
domains. Mean signed bias and RMS error are reported for the total kinetic energy
$e=\frac{1}{3}\sum_\alpha\langle u_\alpha^2\rangle$, together with the maximum
component-wise deviation from the mean bias, which quantifies residual
anisotropy.}
\label{tab:equipartition_3D}
\begin{tabular}{c|c|c|c}
\hline
\hline
Configuration &
$\overline{\Delta e}$ &
$\mathrm{RMS}(e)$ &
$\max_\alpha \big|\overline{\Delta u_\alpha^2}-\overline{\Delta e}\big|$ \\
\hline
C1 & $1.42\times10^{-6}$ & $1.62\times10^{-6}$ & $7.0\times10^{-8}$ \\
C2 & $1.43\times10^{-6}$ & $1.62\times10^{-6}$ & $9.0\times10^{-8}$ \\
C3 & $1.25\times10^{-6}$ & $2.26\times10^{-6}$ & $2.0\times10^{-8}$ \\
C4 & $8.30\times10^{-7}$ & $1.25\times10^{-6}$ & $1.8\times10^{-7}$ \\
\hline
\hline
\end{tabular}
\end{table}
\\
\indent A representative instantaneous snapshot of the fluctuating velocity field is shown in Figure~\ref{Figure7} for each scenario at the end of the simulations. All velocity fields exhibit spatially uncorrelated, noise-like structures with no visible alignment or large-scale organisation, consistent with equilibrium thermal fluctuations. Importantly, no directional bias or imprint of the domain aspect ratio is observed, providing qualitative confirmation of isotropy and homogeneity in three dimensions.

\begin{figure}
    \centering
    \includegraphics[width=0.99\linewidth]{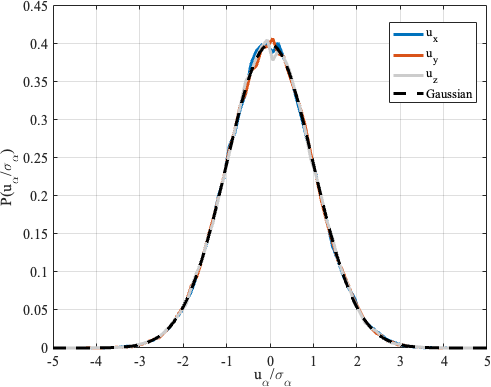}
    \caption{Test 7: scenario C1. Probability density functions of the velocity components at equilibrium, obtained from block histograms of the instantaneous velocity field and normalised by their empirical standard deviations. The distributions of $u_x$, $u_y$, and $u_z$ collapse onto a unit-variance Gaussian, confirming correct thermalisation and isotropy of the fluctuating hydrodynamic modes. Minor apparent offsets of the histogram peaks are attributable to finite binning effects and do not correspond to a measurable bias in the underlying velocity statistics, whose means remain negligible compared to the thermal variance.}
    \label{Figure8}
\end{figure}

Further quantitative insights are given in Figure~\ref{Figure8}, where the probability density functions $P$ of the velocity components (normalised by their variance $\sigma_{\alpha}$) sampled at $t=100 \, t_0$ are plotted for the representative scenario C1. We observe centred at zero and collapse onto a unit-variance Gaussian when normalised by their empirical standard deviations, demonstrating correct thermalisation of the hydrodynamic modes. The excellent agreement across all velocity components confirms isotropy at the level of the full distribution, providing a stringent validation of the fluctuating collision operator beyond second-order moment statistics. These results confirm that the proposed formulation reproduces the full equilibrium statistics of fluctuating hydrodynamics in three dimensions, both at the level of moments and of the complete velocity distribution.

\subsection*{Summary and implications}
Taken together, these seven tests provide a comprehensive validation of the fluctuating lattice Boltzmann implementations. Both BGK and central-moments-based formulations recover the correct equilibrium statistics away from extreme parameter regimes. However, as the relaxation time approaches its lower stability limit, the central-moments formulation exhibits systematically improved stability and statistical consistency. This property is particularly relevant for fluctuating hydrodynamics at low viscosities and high Reynolds numbers, where fluctuating BGK schemes become ill-posed at the discrete level. By ensuring a consistent alignment between dissipation, stochastic forcing, and kinetic mode structure, the present formulation enables stable fluctuating lattice Boltzmann simulations in regimes where classical BGK-based approaches fail.

\section{Conclusions}
In this work, a fluctuating lattice Boltzmann formulation based on orthogonal central moments has been developed in a systematic and physically consistent manner. Thermal fluctuations are introduced directly in central-moment space and paired with mode-dependent relaxation, yielding stochastic collision operators that satisfy the fluctuation--dissipation theorem exactly at the lattice level while preserving mass and momentum conservation.\\
\indent The use of orthogonal central moments plays a pivotal role in the construction. Owing to weighted orthogonality, the equilibrium covariance of the kinetic modes is diagonal, allowing each non-conserved mode to relax independently and to be driven by statistically uncorrelated stochastic forcing. As a result, each kinetic degree of freedom evolves as an independent discrete Ornstein--Uhlenbeck process with variance fixed by equilibrium statistical mechanics, without the need for correlated noise or \textit{ad hoc} regularisation procedures.\\
\indent The numerical tests further clarify the structural nature of this result. In particular, deterministic benchmarks reveal that orthogonal and non-orthogonal central-moment formulations yield identical post-collision populations in the absence of stochastic forcing, despite differing at the level of intermediate moments. This equivalence, however, is specific to the deterministic setting and breaks down once thermal fluctuations are introduced. In non-orthogonal formulations, non-zero equilibrium contributions in higher-order moments necessitate correlated noise to enforce fluctuation--dissipation balance, whereas orthogonal formulations admit diagonal equilibrium covariance and independent stochastic forcing by construction.\\
\indent Explicit fluctuating schemes have been derived for the D2Q9, D3Q19 and D3Q27 lattices, demonstrating that the approach is lattice-agnostic provided that an orthogonal basis of central moments is paired with a Hermite-consistent \textit{complete} equilibrium distribution. Numerical results confirm exact equipartition, correct scaling with thermal energy and density, isotropy of equilibrium fluctuations, and robustness across a wide range of relaxation parameters, including regimes in which fluctuating BGK formulations become ill-posed.\\
\indent The present formulation provides a structurally consistent lattice framework in which deterministic equilibrium structure, dissipation and stochastic forcing are naturally aligned, yielding diagonal equilibrium covariance and statistically independent kinetic modes.

\section*{Supplemental material}
The supplemental material is available at \url{https://github.com/SIG-LBM-Multiphysics-Modelling/Fluctuating-CMs-LBM}. It includes MATLAB scripts for constructing the D2Q9, D3Q19 and D3Q27 central-moment lattice Boltzmann schemes, as well as C++ programs used to reproduce the numerical simulations presented in this work. The C++ implementations are written using the \texttt{Kokkos} library~\cite{9485033}, ensuring performance portability across diverse hardware architectures.

\section*{Acknowledgements}
The authors are grateful to Dr Abinhav Muta for his help with the \texttt{Kokkos} library.

\section*{Declaration of Interests}
The authors report no conflict of interest.

\bibliography{biblio}

\end{document}